\documentclass[aps,pra,reprint,amsmath,amsfonts,nofootinbib]{revtex4-1}

\usepackage[T1]{fontenc}
\usepackage[utf8]{inputenc}
\usepackage{braket}
\usepackage{graphicx}
\usepackage{siunitx}

\newcommand{\effm}{{m^*}}

\newcommand{\conj}{^*}
\newcommand{\transpose}{^\mathrm{T}}

\DeclareMathOperator{\dif}{d \!}

\DeclareMathOperator{\sech}{sech}
\DeclareMathOperator{\csch}{csch}

\DeclareMathOperator{\imag}{Im}

\DeclareMathOperator{\Tr}{Tr}
\DeclareMathOperator{\nullspace}{Null}

\newcommand{\od}[3][]{\ensuremath{%
  \frac{\dif{^{#1}}#2}{\dif{#3^{#1}}}}}
\newcommand{\pd}[3][]{\ensuremath{%
  \frac{\partial{^{#1}}#2}{\partial#3^{#1}}}}
  
\newcommand{\half}{{\ensuremath{\tfrac{1}{2}}}}

\newcommand{\dihedral}[1]{\mathrm{D}_{#1}}

\renewcommand{\vec}[1]{\boldsymbol{#1}}

\newcommand\matlab{\textsc{Matlab}}

\newcommand\kpar{k_{m\parallel}}
\newcommand\kper{k_{m\perp}}

\newcommand{\rleg}{\ensuremath{\mathrm{E}}}
\newcommand{\uleg}{\ensuremath{\mathrm{N}}}
\newcommand{\lleg}{\ensuremath{\mathrm{W}}}
\newcommand{\dleg}{\ensuremath{\mathrm{S}}}

\usepackage{xcolor}

\begin{document}

\title{Physical approach to quantum networks with massive particles}

\author{Molte Emil Strange Andersen}
\affiliation{Department of Physics and Astronomy, Aarhus University,
DK-8000 Aarhus C, Denmark} 

\author{Nikolaj Thomas Zinner}
\affiliation{Aarhus Institute of Advanced Studies, Aarhus University, DK-8000 Aarhus C, Denmark}
\affiliation{Department of Physics and Astronomy, Aarhus University,
DK-8000 Aarhus C, Denmark} 

\date{\today}

\begin{abstract}
Assembling large-scale quantum networks is a key goal of modern physics research with applications in quantum information and computation. Quantum wires and waveguides in which massive particles 
propagate in tailored confinement is one promising platform for realizing a quantum network. 
In the literature, such networks are often treated as quantum graphs, that is, the wave functions are taken to live on graphs of one-dimensional edges meeting in vertices.
Hitherto, it has been unclear what boundary conditions on the vertices produce the physical states one finds in nature.
This paper treats a quantum network from a physical approach, explicitly finds the physical eigenstates and compares them to the quantum-graph description.
The basic building block of a quantum network is an X-shaped potential well made by crossing two quantum wires, and we consider a massive particle in such an X well.
The system is analyzed using a variational method based on an expansion into modes with fast convergence and it provides a 
very clear intuition for the physics of the problem.
The particle is found to have a ground state that is exponentially localized to the center of the X well, and the other symmetric solutions are formed so to be orthogonal to the ground state.
This is in contrast to the predictions of the conventionally used so-called Kirchoff boundary conditions in quantum graph theory that predict a different sequence of symmetric solutions that cannot be physically realized.
Numerical methods have previously been the only source of information on the ground-state wave function and our results  
provide a different perspective with strong analytical insights.
The ground-state wave function has the shape of a solitonic solution to the non-linear Schr{\"o}dinger equation, enabling an analytical prediction of the wave number.
When combining multiple X wells into a network or grid, each site supports a solitonic localized state.
The solitons only couple to each other and are able to jump from one site to another as if they were trapped in a discrete lattice.
\end{abstract}
\maketitle

\tableofcontents

\begin{figure}
  \includegraphics{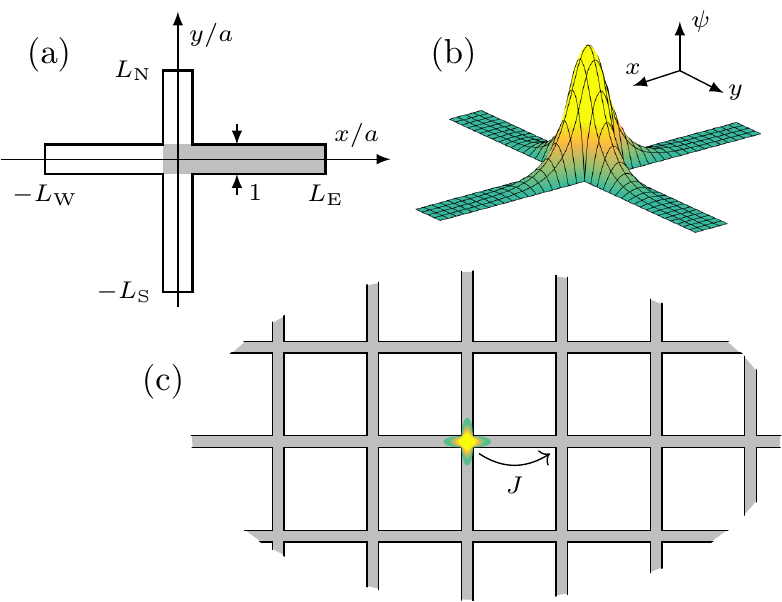}
  \caption{
    The X well constituting part of a quantum network.
    (a)
    Geometry of the two-dimensional X well. The thick boundary marks an infinitely high potential barrier beyond which the wave function must vanish. The legs have lengths $a L_s$, where $s=\rleg$, $\uleg$, $\lleg$, $\dleg$, measured from the origin of the coordinate system. All the legs are of equal width, namely $a$. The shading illustrates the area $A_\rleg = \{ -\frac{1}{2}a \le x \le aL_\rleg, |y| \le \frac{1}{2}a \}$ in which the eastern modes $\ket{m,\rleg}$ reside.
    (b)
    Surface plot of ground-state wave function $\psi(x,y)$ for a symmetric X well with $L_s=5$.
    (c)
    A soliton in a network of X wells.
    The soliton has a probability per unit time $J$ for jumping to a neighboring site.%
  }
  \label{fig:xwell-schematic}
\end{figure}

\section{Introduction}

In the effort towards creating a quantum computer, much attention has been given to designing and producing electronic devices on the quantum scale \cite{ladd2010,kane1998,zwanenburg2013,maurer2012,awschalom2013}.
Only to a lesser extend has one focused on how the devices are connected in circuits of quantum wires or waveguides.
Multiple platforms promise to realize a quantum network; among them are mesoscopic semiconductor devices \cite{timp1988,worschech2001}, carbon nanotubes \cite{papadopoulos2000,terrones2002} and carbon nanowires \cite{mayorov2011,ni2006,li2008,cai2010}.
In order to understand the behavior of massive particles in  a quantum network, it is necessary to first obtain a thorough understanding of the basic building block of such a network, namely, an intersection of quantum wires.
In this paper, we study such a wire crossing and how it forms part of the network.

In order to set the stage for describing the essential components of the network, we take two flat quantum wires and place them perpendicular to one another in the same plane such that they intersect, as illustrated in Fig.~\ref{fig:xwell-schematic}(a).
Inside the wires, we place a massive particle and assume that it is free to move in the wires.
We will denote this setup an \emph{X well}.
We assume the particle is forbidden to move outside the X well, that is, that the potential barriers surrounding the X well are infinitely high.
We imagine that an X well could be built using carbon nanowires, but will not restrict ourselves to a specific realization.
If we are able to describe the physics of a network of wires, one might infer that we are also able to understand the essentials of other networks with different trapping potentials, as for instance those realized for cold atoms using an optical lattice \cite{stockhofe2014,amico2017}. 
The setup is also of interest in classical physics, since solving the Schrödinger equation on an X well is equivalent to finding the eigenmodes of a drum whose membrane has the shape of an `X'.

The X well has been studied in some detail previously \cite{schult1989,avishai1991,markowsky2014,exner1996,delitsyn2012,voo2006,bulgakov2002}, but mainly as an open system and with focus on the ground state.
In this paper, we impose Dirichlet boundary conditions ($\psi=0$ for a wave function $\psi$) at the end walls and consider both the ground state and the excited states.
We solve the problem using a variational method that is essentially similar to the one employed in \textcite{avishai1991}, though our method is a more general formulation able to describe not only the ground state but also excited states.

In the literature, `quantum networks' bear many meanings. An early application of the 
term can be found in \textcite{yurke1984} which deals with low-noise electromechanical 
networks in the quantum regime and discusses the essentials of quantization of electrical 
circuits. More recently, it has become more common to discuss quantum networks in the 
context of hybrid platforms that may help create a {\it quantum internet} \cite{kimble2008}
where quantum entanglement and teleportation is spread across many nodes. Important examples of 
such hybrid structures are photonic crystals and nanostructures \cite{lodahl2015,lodahl2017}, 
cavity-based light-matter systems with atoms and optical photons \cite{ritsch2013,reiserer2015}, 
cavity optomechanics \cite{aspelmeyer2014},
ion traps integrated with photonics \cite{duan2010,monroe2013}, and superconducting circuits 
integrated with microwave cavities \cite{blais2004,girvin2009,xiang2013}. These hybrid quantum 
systems \cite{kurizki2015} are expected to become a backbone for future quantum simulations \cite{georgescu2014}.
In most of these platforms, 
information is conveyed between nodes by photons. This is different from the present context
of a quantum network where the carriers are massive particles. On the other hand, our setup 
is directly relevant for recent work on implementing arbitrary optical wave guides for cold 
atoms using for instance `painted potentials' \cite{ryu2015} with the purpose of producing 
quantum circuits with atoms, so-called `atomtronics' \cite{amico2017}. The basic building 
blocks consisting of cross-beam waveguides have already been experimentally demonstrated
\cite{gattobigio2012,mcdonald2013}.

Our study draws close parallels to the mathematical field of quantum graph theory, which is the theory of differential operators on graphs of one-dimensional edges connected by vertices.
Quantum graphs were first used by \textcite{exnerseba1989} to analyse bound states in bent waveguides.
Since then they have been used extensively to describe variuos phenomena in physics and chemistry.
Apart from the present application with circuits of quantum wires, quantum graph theory are among other things used to model 
solids \cite{richardson1972}, photonic crystals \cite{kuchment2001}, microwaves in waveguides, cavities and resonators \cite{delitsyn2012}, superconductors \cite{alexander1983,rubinstein1997,rubinstein1998,jeffery1989}, atomic and molecular wires \cite{joachim1997}, spin-orbit interactions \cite{harrison2008} and quantum chaos \cite{kottos1997,kottos1999} on networks, and to model aromatic carbohydrate molecules \cite{griffith1952,ruedenberg1953}.
Quantum graphs have been investigated both using differential operators and from a scattering-matrix approach \cite{berkolaiko2008,kostrykin1999}.
Several reviews are available on the topic; see for instance, \textcite{kuchment2002,kuchment2008}.

It is known that for a `fat' quantum graph -- i.e., a graph whose edges have a non-vanishing thickness -- with Dirichlet boundary conditions that collapses into a quantum graph without transverse extension, the resulting effective boundary conditions depend on the geometry of the fat graph \cite{cacciapuoti2007,grieser2008,exner2005,molchanov2007,post2005}.
This is in contrast to a fat graph with Neumann boundary conditions, i.e., where the derivative of the wave function is zero at the boundaries \cite{kuchment2001b,rubinstein2001,saito2000}.
The Dirichlet problem is in general identified in the mathematical literature as being difficult, though some progress has been made in recent years.
One can argue that the X well is the simplest non-trivial example of a fat graph with Dirichlet boundary conditions, so the present study also serves to test the assumptions and claims of quantum graph theory.

In order to rigorously define the intersection or vertex region, we call the intersecting area of the two wires in the X well the \emph{central region} and the rest \emph{legs}.
As shown on Fig.~\ref{fig:xwell-schematic}(a), the width of each wire is $a$ and the length of each leg as measured from the center of the well is $aL_s$, where $s=\rleg$ (east), $\uleg$ (north), $\lleg$ (west) or $\dleg$ (south) denotes the leg in question.
The main priority is now to 
find the energy eigenstates of a particle with (effective) mass $\effm$ in the wires of the X well.
This amounts to solving the Schrödinger equation
\begin{equation}
  -\frac{\hbar^2}{2\effm} \nabla^2 \psi + V \psi = E \psi.
\end{equation}
Here, $\nabla^2$ is the two-dimensional Laplace operator.
Inside the X well, the particle is free to move and the Schrödinger equation is equivalent to the two-dimensional Helmholtz equation
\begin{equation}
  (k^2 + \nabla^2) \psi = 0, \label{eq:helmholtz}
\end{equation}
where $k$ is the wave number, defined through $E = \hbar^2k^2/2\effm$. Throughout our discussion, we will assume that the system is at zero temperature in order to ensure that our network is in the fully quantum regime, i.e. that the temperature scale is below the energy gap between ground and first excited states in the system. Experimental access to this regime has been demonstrated in many of the platforms discussed above.

The outline of this paper is as follows. In Sections~\ref{sec:xwell-modes} to \ref{sec:symmetric-xwell} we solve the Helmholtz equation and discuss the properties of the solutions.
We find that the ground state of the particle -- as plotted in Fig.~\ref{fig:xwell-schematic}(b) -- is exponentially localized to the center of the X well.
Due to this exponential behavior of the wave function, the lengths of the legs do not have appreciable impact on the ground-state wave function (if beyond a certain size).
Neither does the boundary conditions on the end walls of the legs -- we show this explicitly for several examples later in the paper.

Interestingly, we find that the cross section of the ground-state wave function has the shape of a solitonic solution to the non-linear Schrödinger equation, namely a hyperbolic secant function. This enables an analytical prediction of the wave number $k\simeq  \sqrt{{2}/{3}}\,{\pi}/{a}$ within $1\%$ of the numerically attained value.

Next, we study the excited states. 
For a symmetric X well, the solutions are characterized in terms of symmetries.
They are found to resemble solutions to the well-known problem of a particle in a box, and the eigenenergies of the former converge to those of the latter for increasing $L_s$.
We apply the solutions in Section~\ref{sec:xwell-wave-propagation} to study how an incoming signal propagates through the X well and find that it is the excited states that determines the transmission properties when the incoming wave has an energy above the transverse excitation threshold.

If the X well is constructed with very thin legs, the naive expectation is that excitations transverse to the legs are inaccessible at low temperatures and that the transverse degree of freedom can be integrated out to obtain an effective one-dimensional description.
This approach -- which we problematise in Section~\ref{sec:quantum-graph-theory} -- results in the X well reducing to a quantum graph with so-called Kirchoff boundary conditions at the center vertex.
These boundary conditions have been routinely employed in theories on quantum networks since the early 1950's \cite{ruedenberg1953,griffith1952} and are sometimes in the literature uncritically assumed to hold \cite{voo2006}.%
\footnote{\textcite{kuhn1954} quickly realized, however, that the Kirchoff boundary conditions may be generalized while still satisfying conservation of probability current. See also the discussions in \textcite{frost1955} and \textcite{kuhn1956}.}
As it turns out, however, the Kirchoff boundary conditions are unable to account for the symmetric solutions to the X well (including the ground state), and they are not the correct boundary conditions for the physical problem at hand.
It has therefore been unclear what boundary conditions -- if any -- correctly reproduces the eigenstates of the X well in the context of quantum graph theory.

We find that for all excited states, the wave function diminishes at the X-well center as the length of the legs are increased relative to their width.
This translates to a simple boundary condition for the associated quantum graph problem, namely that the wave function must vanish at the graph vertex.
The decoupling of the X-well legs thus achieved hinders the propagation of waves through the network.
As the legs connecting the X'es in a physical network, however, must be of finite length, the trivial dynamics predicted by quantum graph theory are not applicable.

Finally, Sections~\ref{sec:xwell-variations} and \ref{sec:xwell-network} study different variations on the X well and considers the X well as an element of a network of quantum wires.
When combining multiple X wells into a network or grid, each site supports a solitonic localized state.
We show that the solitons only couple to each other and not to other classes of states in the spectrum. Furthermore,
they are able to jump from one site to another, and the X-well network is, thus, a realization of a lattice -- 
see Fig.~\ref{fig:xwell-schematic}(c). This emerging lattice of localized solutions may be interesting 
for precision sensing and metrology \cite{degen2017} due to their diminshing coupling 
to other states and thus increased robustness.

\section{Modes as solutions to the Schrödinger equation}
\label{sec:xwell-modes}

The geometry of the problem does not allow one to separate the spatial coordinates $x$ and $y$.
We can, however, partition the X well into four (overlapping) rectangular regions $A_s$ in which separation of variables may be employed, as detailed below.
The rectangles are chosen such that they each encompass the central region together with one leg.
This is exemplified for $s=\rleg$ as the shaded area in Fig.~\ref{fig:xwell-schematic}(a).

We solve Eq.~\eqref{eq:helmholtz} in each of the rectangles $A_s$ and take the solution to vanish outside the given rectangle.
We shall call such a solution a \emph{mode}.
Contrary to how the word is sometimes used in the literature, in our context a single mode is not an eigenstate of the system.
An energy eigenstate is a superposition of modes.

The following two-step procedure is used to construct the mode wave functions.
Throughout, we shall illustrate the steps in the procedure with the eastern mode $A_\rleg$ as an example.

\paragraph*{Step 1.}
Take the ansatz for a wave function
\begin{equation}
  \psi(x,y) = X(x) Y(y),
\end{equation}
for some functions (of a single variable) $X$ and $Y$ describing the mode wave function along the leg and transverse to it.
Inserting the ansatz into the Helmholtz equation gives us
\begin{equation}
  \od[2]{}{x}X(x) = -k_\parallel^2 X(x), \quad\text{and}\quad
  \od[2]{}{y}Y(y) = -k_\perp^2 Y(y), \label{eq:helmholtz-separated}
\end{equation}
for constants $k_\perp$ and $k_\parallel$ subject to the constraint $k^2 = k_\perp^2 + k_\parallel^2$.

To ensure that the mode wave function vanishes at the boundaries above and below the leg, i.e., at $\{\half a\le x\le aL_\rleg, |y|=\half a\}$, we take the transverse solution to be
\begin{equation}
  Y(y) = \sin(\pi m(\half + y/a)) \label{eq:transverse-mode-wave-function}
\end{equation}
for a positive integer $m$.
We label the mode by its quantum number $m$ and denote it in Dirac notation $\ket{m,\rleg}$.
With the solution Eq.~\eqref{eq:transverse-mode-wave-function}, we have set the transverse component of the wave vector to $k_{m\perp}=m\pi/a$. Notice that since we take $a$ as the basic unit of length, the length parameters for the legs, $L_s$, are dimensionless. 

As the longitudinal mode we could take a solution of the form $X(x) = \sin(\frac{n\pi}{L_\rleg} (\half + y/a))$, but this is too restrictive since it fixes the energy completely.

\paragraph*{Step 2.}
The rectangle $A_s$ is divided into two disjoint sub-rectangles $A^{\text{int}}$ and $A_s^{\text{ext}}$.
One of them constitutes the central region, $A^{\text{int}} = \{ |x|,|y| \le \frac{1}{2}a \}$, which is the same for all $s$.
The other rectangle, $A_s^{\text{ext}}$, is a leg, e.g., 
\begin{equation}
  A_\rleg^{\text{ext}} = \{ \half a\le x\le aL_\rleg, |y| \le \half a \}.
\end{equation}

While the transverse part of the mode wave function ($Y(y)$ in our example) is the same for $A^\text{int}$ and $A_s^\text{ext}$, the longitudinal part ($X(x)$) is different.
We normalise the mode such that the longitudinal part of its wave function is $1$ at the interface between the central region $A^{\text{int}}$ and the leg $A_s^{\text{ext}}$.
(Note that this implies that $\braket{m,s|m,s}\not=1$.)
We require the wave function to be continuous.

For the eastern mode, the wave function $\psi(x,y)=\braket{x,y|m,\rleg}$ is
\begin{equation}
  \braket{x,y|m,\rleg} = \csc(k_{m \parallel} a) 
  \sin(k_{m \parallel}(\tfrac{1}{2}a + x)) \sin(k_{m\perp} (\tfrac{1}{2}a + y))
\end{equation}
for $(x,y) \in A^{\text{int}}$, i.e., in the central region.
Meanwhile
\begin{multline}
  \braket{x,y|m,\rleg} = \csc(k_{m \parallel} a(L_\rleg - \tfrac{1}{2}))  \\
  \cdot \sin(k_{m \parallel} (a L_\rleg - x)) \sin(k_{m\perp} (\tfrac{1}{2}a + y))
\end{multline}
for $(x,y) \in A_s^{\text{ext}}$, i.e., in the leg.
This construction ensures that $\braket{-\half a,y|m,\rleg}=\braket{L_\rleg,y|m,\rleg}=0$ as required and that $\braket{\half a,y|m,\rleg}=\sin(k_{m\perp} (\tfrac{1}{2}a + y))$.
However, the derivative of the mode wave function is, unfortunately, discontinuous at the boundaries of $A^{\text{int}}$ and $A_s^{\text{ext}}$ (including the interface between them).

Notice at this point that the longitudinal component of the wave vector, $k_{m\parallel}$, may be imaginary, in which case $k < k_{m\perp}$.
In this case, the trigonometric functions of $k_{m\parallel}$ turn into hyperbolic functions, and the mode wave function has an exponential behavior.
As we shall see later, this turns out to be critically important for the description of the ground state in the X well.

\section{Eigenstates as mode expansions}
\label{sec:xwell:eigenstates-as-mode-expansions}

Having obtained a complete set of modes using the above-described procedure, we turn towards finding the energy eigenstates of the full system (the entire X well).
We can write an energy eigenstate $\ket\psi$ as a linear combination of modes:
\begin{equation}
  \ket{\psi} = \sum_{m=1}^\infty \sum_{s=\rleg,\uleg,\lleg,\dleg} \alpha_{ms} \ket{m,s}. \label{eq:infinite-mode-sum}
\end{equation}
The coefficients $\alpha_{ms}$ must be chosen such that the wave function $\psi(x,y) = \braket{x,y|\psi}$ is continuously differentiable within the X well.

Alternatively, this can be stated as a variational principle; the coefficients $\alpha_{ms}$ must be chosen among the stationary points of the energy functional
\begin{equation}
  E[\ket{\psi}] = \frac{\hbar^2}{2\effm} \left( k^2 - \frac{\braket{\psi|\Pi|\psi}}{\braket{\psi|\psi}} \right). \label{eq:energy-functional}
\end{equation}
Here, $\Pi$ is the operator giving the energy contribution due to the kinks in the mode wave functions at the interfaces between the legs and the central region.
Its expectation value is
\begin{align}
  \braket{\psi|\Pi|\psi}
  &= \lim_{\epsilon\downarrow0} \int_{-\half a}^{\half a} \dif{y} \; \int_{-\half a - \epsilon}^{\half a + \epsilon} \dif{x} \; \psi\conj(x,y) \pd[2]{\psi}{x} + \dotsb \notag \\
  &= \int_{-\half a}^{\half a} \dif{y} \; \psi\conj(\half,y) \; \Delta\!\left( \pd{\psi}{x} \right)_{x=\half a} + \dotsb,
\end{align}
where $\Delta$ denotes the change when crossing the interface and `$\dotsb$' stands for the similar contributions from the $\uleg$, $\lleg$ and $\dleg$ interfaces.

We are primarily interested in the ground state and the first few excited states.
Because we have constructed the modes such that the $m$'th mode corresponds to a transverse excitation with energy $\frac{\hbar^2\pi^2}{2\effm a^2} m^2$,
in the limit $L \gg 1$ and at low temperatures only the lower modes of the infinite sum in Eq.~\eqref{eq:infinite-mode-sum} contribute significantly because transverse excitations of the legs require a lot of energy.
Motivated by this consideration, we may, as a trial state, take a sum of a finite subset of the terms in Eq.~\eqref{eq:infinite-mode-sum} and use this as the basis in a variational calculation.

The energy of a mode expansion in a \emph{finite} number of modes $\{\ket{m,s}\}$ with wave number $k$ and mode coefficients given by the vector $\vec{\alpha}$, may be stated as
\begin{equation}
  E(k,\vec{\alpha}) = \frac{\hbar^2}{2\effm} \left( k^2 - \frac{\vec{\alpha}^\dagger \Pi \vec{\alpha}}{\vec{\alpha}^\dagger \Psi \vec{\alpha}} \right)
\end{equation}
where $\Pi$ is the matrix of elements $\braket{m,s|\Pi|m',s'}$, and $\Psi$ is the matrix of overlaps $\braket{m,s|m',s'}$ between modes $\ket{m,s}$ and $\ket{m',s'}$.

\subsection{Finding the ground state by variation}

We are now ready to formulate a numerical method to find the eigenstates.
We start with the ground state.

To find a mode expansion that is close to the ground state, we minimise the energy $E(k,\vec{\alpha})$,
\begin{align}
  \min_{k,\vec{\alpha}} E(k,\vec{\alpha})
  &= \min_k \frac{\hbar^2}{2\effm} \left( k^2 - \max_{\vec{\alpha}} \frac{\vec{\alpha}^\dagger \Pi(k) \vec{\alpha}}{\vec{\alpha}^\dagger \Psi(k) \vec{\alpha}} \right) \notag \\
  &= \min_k \frac{\hbar^2}{2\effm} \left( k^2 - \lambda_\text{max}(k) \right), \label{eq:minE}
\end{align}
where $\lambda_\text{max} = \max \{ \lambda \;\vert\! \det(\Pi - \lambda \Psi) = 0 \}$ is the greatest generalized eigenvalue of $\Pi$ with respect to $\Psi$ for a given wave number $k$.
Numerically, we may compute Eq.~\eqref{eq:minE} by a one-dimensional downhill simplex routine by which the optimum wave number is located.
Once the ground-state wave number $k_\text{gs}$ has been found, the mode coefficients $\vec{\alpha}$ are found as the generalized eigenvector corresponding to the eigenvalue $\lambda_\text{max}(k_\text{gs})$.

As an alternative to computing the matrix elements of $\Pi$ (in order to find its eigenvalues), we may perform an integration by parts of the Laplacian in Eq.~\eqref{eq:helmholtz}:
\begin{align}
  \braket{\psi|H|\psi}
  &= -\frac{\hbar^2}{2\effm} \int \dif{x}\,\dif{y}\; \psi\conj \nabla^2\psi \notag \\
  &= \frac{\hbar^2}{2\effm} \int \dif{x}\,\dif{y}\; (\nabla\psi\conj) \cdot (\nabla\psi). \label{eq:H-partial-int}
\end{align}
Whereas kinks in the wave function $\psi$ give rise to energy contributions through Dirac-delta terms in $\nabla^2\psi$, the gradient $\nabla\psi$ is discontinuous, but integrable.
So when using Eq.~\eqref{eq:H-partial-int} to compute the matrix elements of the Hamiltonian, one does not have to worry about extra terms arising from the kinks in $\psi$.
With this alternate approach, we would compute the matrix elements of $H$ from Eq.~\eqref{eq:H-partial-int} and find the \emph{minimal} generalized eigenvalue of $H$ with respect to $\Psi$ under variation of $k$.

\subsection{Excited states}
\label{sec:xwell-excited-states}

In the preceding section, we described how to find the ground state.
We now proceed to the excited states.
The excited states are found through an orthogonalization procedure:
Assume we have already found the lowest $N$ eigenstates and denote their wave numbers $k_1,\dotsc,k_N$ and their coefficient vectors $\vec{\alpha}_1,\dotsc,\vec{\alpha}_N$.
Then by orthogonality of eigenstates, we must seek the coefficient vector $\vec{\alpha}_{N+1}$ of the $(N+1)$'th eigenstate among the vectors satisfying $\vec{\alpha}_i^\dagger \Psi(k_i,k_{N+1}) \vec{\alpha}_{N+1} = 0$ for all $i=1,\dotsc,N$.
(Here, $\Psi(k_i,k_{N+1})$ denotes the matrix of overlaps between modes of different energy.)
Thus,
\begin{equation}
  \vec{\alpha}_{N+1} \in \nullspace\!\begin{bmatrix}
    \vec{\alpha}_1^\dagger \Psi(k_1,k_{N+1}) \\
    \vdots \\
    \vec{\alpha}_N^\dagger \Psi(k_N,k_{N+1})
  \end{bmatrix}. \label{eq:orthogonality-condition}
\end{equation}

If $B$ is a matrix whose columns span the null space above, we find the desired eigenstate by solving the generalized eigenvalue problem
\begin{equation}
  (B^\dagger \Pi B) \vec{\beta} = \tilde{\lambda} (B^\dagger \Psi(k_{N+1}) B) \vec{\beta},
\end{equation}
where $\vec{\alpha}_{N+1} = B\vec{\beta}$ and $\tilde{\lambda}$ is the eigenvalue.
Let $\tilde{\lambda}_\text{max}(k)$ be the greatest among the generalized eigenvalues at a given wave number $k$.
We use the same minimization procedure as for the ground state, that is, finding the wave number $k_{N+1}$ that minimizes
\begin{align}
  E(k)
  &= \frac{\hbar^2}{2\effm} \left( k^2 - \max_{\vec{\beta}} \frac{\vec{\beta}^\dagger B^\dagger \Pi(k) B \vec{\beta}}{\vec{\beta}^\dagger B^\dagger \Psi(k) B \vec{\beta}} \right) \notag \\
  &= \frac{\hbar^2}{2\effm} \left( k^2 - \tilde{\lambda}_\text{max}(k) \right).
\end{align}

Notice that since, in practice, the lower eigenstates are only approximately known, variational bounds cannot be guaranteed for excited states \cite{bransden-joachain}.

\section{The symmetric X well}
\label{sec:symmetric-xwell}

\begin{figure}
  \centering
  \includegraphics{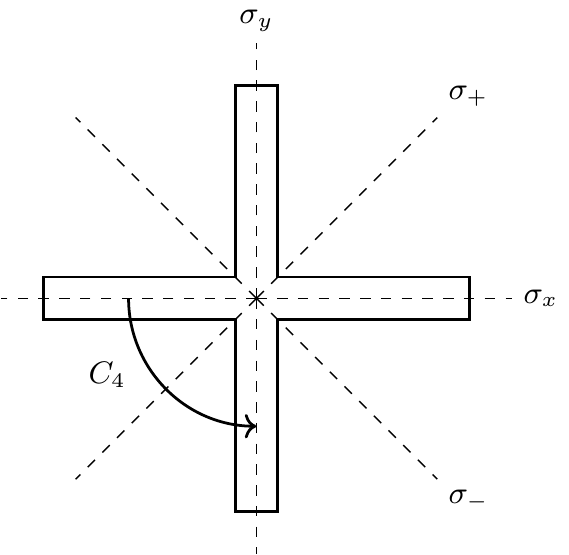}
  \caption{The symmetric X well marked with its four reflection axes (dashed lines) and the discrete, rotational symmetry.}%
  \label{fig:xwell-symmetries}
\end{figure}

\begin{table}
  \centering
  \begin{ruledtabular}
    \begin{tabular}{cll}
      $E$ & $(x,y)\mapsto (x,y)$ & The identity element \\
      $C_4$ & $(x,y)\mapsto (-y,x)$ & Rotation by $\pi/2$ \\
      $C_4^2$ & $(x,y)\mapsto (-x,-y)$ & Rotation by $\pi$ (parity) \\
      $C_4^3$ & $(x,y)\mapsto (y,-x)$ & Rotation by $-\pi/2$ \\
      $\sigma_+$ & $(x,y)\mapsto (y,x)$ & Reflection in the line $y=x$ \\
      $\sigma_-$ & $(x,y)\mapsto (-y,-x)$ & Reflection in the line $y=-x$ \\
      $\sigma_x$ & $(x,y)\mapsto (x,-y)$ & Reflection in the $x$-axis \\
      $\sigma_y$ & $(x,y)\mapsto (-x,y)$ & Reflection in the $y$-axis \\
    \end{tabular}
  \end{ruledtabular}
  \caption{The elements of the dihedral group $\dihedral4$ and their respective actions as representations on wave functions.}
  \label{tbl:D4-elements}
\end{table}

In the following we restrict our attention to the symmetric X well, whose legs are all of equal length, $L_s=L$.
In this case, the geometry possesses rotation and reflection symmetries as drawn in Fig.~\ref{fig:xwell-symmetries}.
The symmetries simplify the calculation and classification of the eigenstates.
In this Section, we find and plot the eigenstates of the symmetric X well.

The symmetric X well has the same symmetries as a square; its symmetry group is the dihedral group of order 8, which is conventionally denoted $\dihedral4$.
The eight operations under which the geometry of the symmetric X well is invariant are outlined in Table~\ref{tbl:D4-elements}.
The group $\dihedral4$ is non-Abelian, so we cannot diagonalize the representations of all group elements simultaneously.
We choose to diagonalize the reflection operators $\sigma_+$ and $\sigma_-$ and classify the energy eigenstates according to their eigenvalues $r_+$ and $r_-$ with respect to $\sigma_+$ and $\sigma_-$, respectively.
If $\braket{\sigma_+}=\{E,\sigma_+\}$ denotes the subgroup generated by $\sigma_+$, and similarly $\braket{\sigma_-}=\{E,\sigma_-\}$, then our classification gives eigenstates that are irreducible representations of the product subgroup
\begin{equation}
  \braket{\sigma_+} \times \braket{\sigma_-} = \{ E, \sigma_+, \sigma_-, C_4^2 \},
\end{equation}
which is isomorphic to the Klein four-group $\mathbb{Z}_2 \times \mathbb{Z}_2$.
The simple characters of the subgroup $\braket{\sigma_+} \times \braket{\sigma_-}$ correspond to compound characters of $\dihedral4$.
Table~\ref{tbl:D4-char-table} lists the simple characters of $\dihedral4$ \cite{hamermesh}.
As the table indicates, the dihedral group $\dihedral4$ has four one-dimensional irreducible representations ($A_1$, $A_2$, $B_1$ and $B_2$ in Mulliken symbols) and a single two-dimensional irreducible representation ($E$).

\begin{table}
  $$\begin{array}{ccccccl}
    \hline
    \mathrm{D}_4 & E & C_4^2 & C_4, C_4^3 & \sigma_+, \sigma_- & \sigma_x,\sigma_y \\
    \hline
    A_1 & \phantom{-}1 & \phantom{-}1 & \phantom{-}1 & \phantom{-}1 & \phantom{-}1 \\
    A_2 & \phantom{-}1 & \phantom{-}1 & \phantom{-}1 & -1 & -1 \\
    B_1 & \phantom{-}1 & \phantom{-}1 & -1 & -1 & \phantom{-}1 \\
    B_2 & \phantom{-}1 & \phantom{-}1 & -1 & \phantom{-}1 & -1 \\
    E & \phantom{-}2 & -2 & \phantom{-}0 & \phantom{-}0 & \phantom{-}0 \\
    \hline
  \end{array}$$
  \caption{%
    Character table for the dihedral group $\dihedral4$.
    When the header entry for a column contains multiple group elements, they belong to the same conjugacy class.%
  }
  \label{tbl:D4-char-table}
\end{table}

Any state with $(r_+,r_-)=(+,-)$ has a degenerate partner state with $(r_+,r_-)=(-,+)$ that is identical to the $(+,-)$ state up to a rotation of the coordinate system by $\pi/2$.
These states belong to the two-dimensional representation $E$.
The character table reveals that the classification with the quantum numbers $(r_+,r_-)$ is insufficient to tell the one-dimensional representations $A_1$ and $B_2$ apart; neither can it distinguish $A_2$ from $B_1$.

As $C_4^2=\sigma_+\sigma_-$, the members of the irreducible representations of $\braket{\sigma_+}\times\braket{\sigma_-}$ are also eigenstates of the parity operator $C_4^2$ with eigenvalue $p=r_{+}r_{-}$.
States with even parity, $p=+1$, are also eigenstates of reflection in the $x$- and $y$-axes, with eigenvalues $r_x=r_y$.
If $r_x=+1$, only modes with odd $m$ are present in the mode expansion, while only even $m$ modes are present if $r_x=-1$.
For this reason, we do not expect states belonging to the $A_2$ and $B_2$ representations at low energies, since their lowest-energy mode has $m=2$, and thus, $\kper=2\pi/a$, while the lowest energy eigenstates in the spectrum have wave numbers $k\simeq\pi/a$.
So even though our classification in $(r_+,r_-)$ quantum numbers is unable to distinguish $A_1$ and $B_1$ from $A_2$ and $B_2$, this is inessential as long as we only consider states with sufficiently low energy that the latter representations do not occur.

\subsection{Mode wave functions}

In this short section, we find the mode wave functions for the symmetric X well.
The eastern, northern, western and southern modes are related as
\begin{equation}
  \begin{gathered}
    \ket{m,\uleg} = \sigma_+ \ket{m,\rleg}, \quad
    \ket{m,\lleg} = C_4^2 \ket{m,\rleg}, \\
    \ket{m,\dleg} = \sigma_- \ket{m,\rleg}.
  \end{gathered}\label{eq:mode-relations}
\end{equation}
Starting with the central wave function of the eastern mode (as found using the procedure given in Section~\ref{sec:xwell-modes})
\begin{equation}
  \braket{x,y|m,\rleg} = \csc(\kpar a) \sin(\kpar(\half a+x)) \sin(\kper(\half a+y)),
\end{equation}
we can then use Table~\ref{tbl:D4-elements} to find the other central wave functions:
\begin{widetext}
\begin{align}
  \braket{x,y|m,\uleg} &= r_+ \csc(\kpar a) \sin(\kpar(\half a+y)) \sin(\kper(\half a+x)), \\
  \braket{x,y|m,\lleg} &= r_+ r_- \csc(\kpar a) \sin(\kpar(\half a-x)) \sin(\kper(\half a-y)), \\
  \braket{x,y|m,\dleg} &= r_- \csc(\kpar a) \sin(\kpar(\half a-y)) \sin(\kper(\half a-x)).
\end{align}
The leg wave functions of the modes can be found similarly from
\begin{equation}
  \braket{x,y|m,\rleg} = \csc(\kpar a(L-\half)) \sin(\kpar(L a-x)) \sin(\kper(\half a+y)).
\end{equation}
\end{widetext}
The matrix elements of $\Psi$ and $\Pi$ are computed in the Appendix.

\subsection{Continuous-derivative approach}

In the following sections, we present two different methods to find the energy eigenstates.
The first method computes the eigenstates by explicitly enforcing a continuous derivative of the wave function at every interface.
The second method -- to which we shall return in Section~\ref{sec:symmetric-xwell-var-ground-state} -- is a variant of the previously described variational method.

The wave function of a single mode is not continuously differentiable across an interface, neither is the sum of all modes with a given $m$ quantum number.
The total wave function $\psi=\sum_{m,s} \alpha_{m} \braket{x,y|m,s}$, however, has to be in order to be an eigenstate.
By symmetry, it suffices to consider only a single interface; we take the interface between the central region and the eastern leg, where $x=\half a$.
Thus, the change in derivative across the interface must be
\begin{equation}
  \Delta\!\left(\pd{\psi}{x}\right)_{x=\half a} = 0.
\end{equation}

Multiply by $\sin(k_{n\perp}(\half a+y))$ for some positive integer $n$ and integrate the $y$-coordinate out to obtain
\begin{multline}
  0 = \sum_{m} \alpha_{m} \int_{-\half a}^{\half a} \dif{y}\; \sin(k_{n\perp}(\half a+y)) \\
  \cdot \Delta\!\left(\pd{}{x}\braket{x,y|m,\rleg}\right)_{x=\half a},
\end{multline}
or equivalently, $\braket{n,\rleg|\Pi|\psi}=0$.

Numerically, we truncate $\Pi$ to have a finite size by choosing a maximum mode number $M$ such that only modes $n,m\le M$ are included.
We then have the $M$-by-$M$ matrix equation $\Pi \vec{\alpha} = \vec{0}$ and we see that the desired mode coefficients $\vec{\alpha}$ must belong to the null space of $\Pi$, and as such, $\Pi$ must be singular.
The task is, thus, to find a wave number $k$ for which $\det(\Pi(k))=0$.

\begin{figure}
  \includegraphics{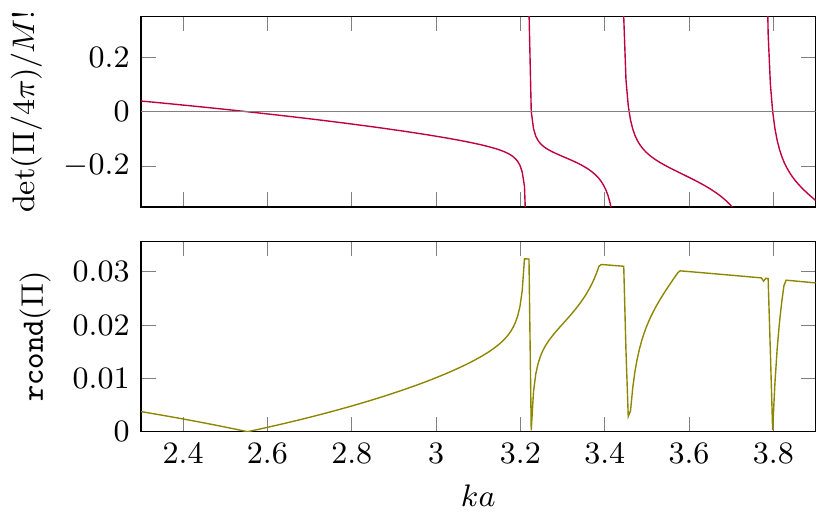}
  \caption{
    The upper panel shows $\det(\Pi)$ for the matrix $\Pi$ for $(r_+,r_-)=(+,+)$, length $L=5$ and cutoff $M=30$.
    $\Pi$ is given by Eq.~\eqref{eq:derivative-matrix-elements}.
    For large $m$, $|t_m|\sim m\pi$, and the diagonal elements of $\Pi$ scales as $4\pi m$, cf.~Eq.~\eqref{eq:Pi-matrix-elements}.
    A determinant scales as $\det(c\Pi)=c^M \det(\Pi)$, so when increasing $M$, the determinant scales rapidly.
    To circumvent this, we downscale the determinant by $\prod_m 4\pi m=(4\pi)^M M!$ before plotting it.
    The lower panel shows the \matlab\ function \texttt{rcond} that is an estimate of the reciprocal condition number of $\Pi$ in $1$-norm.
    The reciprocal condition number is small at the positions of the eigenstate wave numbers -- notice how the dips in \texttt{rcond} correspond to zeros in $\det(\Pi)$.}
    \label{fig:singularities}
\end{figure}

It is straightforward to plot $\det(\Pi)$ as a function of $k$ and get a rough idea about the location of the roots; see Fig.~\ref{fig:singularities}.
It turns out, however, that for the excited states, the determinant crosses zero very rapidly, going from a very high value to a very low, negative value (or vice versa).
A double-precision number can, therefore, be insufficient to resolve the precise location of the root.
Still, the wave number $k$ found with this method may be used as starting point for a finer search with a different method.

Another measure of the `closeness' of a matrix to being singular is the so-called condition number.
The $p$-norm condition number of a matrix $A$ is typically defined as $\lVert A \lVert_p \,\cdot\, \lVert A^{-1} \lVert_p$. 
A large condition number indicates that the matrix is close to being singular.
The condition number may be used in a minimization routine to find the wave number $k$ of an eigenstate as is indicated in the lower panel of Fig.~\ref{fig:singularities}.

Due to the finite size of $\Pi$ and numerical limitations, $\Pi(k)$ might not be exactly singular (that is, $\det(\Pi)$ is small but non-zero) at the wave number of an eigenstate $k$.
If this is the case, the null space of $\Pi(k)$ is trivial, $\nullspace \Pi(k)=\{\vec{0}\}$, and we are, a priori, unable to find the mode coefficients $\vec{\alpha}$.
To circumvent this problem, we can compute the null space of a singular matrix that is close to $\Pi(k)$.

Assume the singular-value decomposition of $\Pi$ is $\Pi=U \Sigma V\transpose$, where $U=[\vec{u}_1,\vec{u}_2,\dotsc,\vec{u}_M]$ and $V=[\vec{v}_1,\vec{v}_2,\dotsc,\vec{v}_M]$ are orthogonal matrices and $\Sigma=\mathrm{diag}(\sigma_1,\sigma_2,\dotsc,\sigma_M)$ is a diagonal matrix containing the singular values of $\Pi$ in descending order.
Then by the Eckart-Young-Mirsky theorem \cite{eckart1936,mirsky1960}, the $M$-by-$M$ matrix
\begin{equation}
  \Pi' = [\vec{u}_1,\dotsc,\vec{u}_{M-1}] \begin{bmatrix}
    \sigma_1 & 0 & \dots & 0 \\
    0 & \sigma_2 & \dots & 0 \\
    \vdots & \vdots & \ddots & \vdots \\
    0 & 0 & \dots & \sigma_{M-1}
  \end{bmatrix} [\vec{v}_1,\dotsc,\vec{v}_{M-1}]\transpose
\end{equation}
is the closest matrix to $\Pi$ in Frobenius norm%
\footnote{The Frobenius norm of a matrix $A$ is defined as $\lVert A \rVert = \sqrt{\Tr(A^\dagger A)}$.}
with rank less than $M$.

We confirm that $\Pi'$ is clearly singular as $\Pi' \vec{v}_M = \vec{0}$ by the orthogonality of $V$.
Hence also, $\vec{v}_M \in \nullspace (\Pi')$, and we may use $\vec{v}_M$ as a basis vector for the approximate null space of $\Pi$ and set $\vec{\alpha}=\vec{v}_M$.
We do not have to consider multi-dimensional null spaces since there is no degeneracy when simultaneously diagonalising $H$, $\sigma_+$, and $\sigma_-$. This is one of the important places at which one sees the power of 
using the group symmetry classification of states discussed above.

\subsection{Variational approach: Ground state}
\label{sec:symmetric-xwell-var-ground-state}

The second method to find the energy eigenstates of the symmetric X well is essentially the variational method described in Section~\ref{sec:xwell:eigenstates-as-mode-expansions}.
The difference with respect to the before-described general case is that the symmetry puts constraints on the modes such that the $\uleg$, $\lleg$ and $\dleg$ modes are given by the $\rleg$ modes, cf.~Eq.~\eqref{eq:mode-relations}.
Therefore, the size of the matrices $\Pi$ and $\Psi$ is reduced from $(4M)\times(4M)$ to $M\times M$, and the computation time required by each diagonalization is reduced by about a factor of $4^3=64$.
More importantly, enforcing a particular symmetry $(r_+,r_-)$ decouples the irreducible representations of $\braket{\sigma_+} \times \braket{\sigma_-}$.
Each family $(r_+,r_-)$ has its own `ground state', and there is no degeneracy in the spectrum that could otherwise be difficult to handle numerically.

Only the $(r_{+},r_{-})=(+,+)$ states are not required to have a node in their wave function at $(x,y)=(0,0)$, and as such, the ground state is expected to belong to the symmetric representation $A_1$.
Using the matrix elements from the Appendix in a numeric implementation of Eq.~\eqref{eq:minE} in \matlab, we find for $L=5$ a ground state with wave number $k = 0.8122\pi/a$.
The wave function is plotted in Fig.~\ref{fig:xwell-schematic}(b).
The ground state is localized about the center of the well; its wave function peaks at the center of the well and decays exponentially along the legs \cite{schult1989}.
Due to this exponential decay, the positions of the end walls are not very important, and the wave function is almost independent of $L$.
The difference in magnitude between the ground-state wave function for $L=3$ and $L=30$ is everywhere less than $0.01/a^2$.
The wave-function overlap between them is 1 up to a tiny deviation of $\sim 10^{-5}$.

The ground-state energy is so small that $k < \pi/a \le \kper$ for every $m$, and thus, $\kpar$ is purely imaginary for all modes.
Define $\zeta_m>0$ such that $k_{m\parallel} = i \zeta_m/a$.
The longitudinal part of the mode wave functions can be rewritten in terms of hyperbolic functions in $\zeta_m$, and this is the cause of the exponential decay along the legs.
Along the eastern leg, for instance,
\begin{equation}
  \braket{x,y|m,\rleg} = \frac{\sinh(\zeta_m (L - x/a))}{\sinh(\zeta_m (L - \tfrac{1}{2}))} \sin(\kper (\half a + y)). \label{eq:eastern-gs-mode}
\end{equation}
The ground state is almost described completely by the first mode.
However, the other modes are required to make the wave function continuously differentiable.
Notice also that since the ground state is almost independent of the lengths of the legs, it is practically the same for the asymmetric X well, as long as the legs are long enough to `saturate' the exponential decay.

\subsection{Soliton approximation}
\label{sec:soliton-approximation}

Perhaps the simplest single-parameter function one can come up with that has a smooth maximum at $x=0$ and falls off exponentially for $|x| \gg a$ is a hyperbolic secant.
Thus, if we wish to describe the $L\to\infty$ limit of the ground state, we may attempt to fit the wave function $\psi_0(x,0) = \sum_{m,s} \alpha_{m} \braket{x,0|m,s}$ to a function of the form $f(x) = A\sech(b x)$, where $A$ and $b$ are positive constants.
We call $f$ a `soliton' because non-linear wave equations like the Gross-Pitaevskii equation support solitonic solutions of this form \cite{kivshar1989,pethick-smith,kartashov2011}. 

For $L\to\infty$, Eq.~\eqref{eq:eastern-gs-mode} turns into
\begin{equation}
  \braket{x,y|m,\rleg} \to e^{-\zeta_m(x/a-\half)} \sin(\kper(\half a+y)).
\end{equation}
Since $\zeta_1 < \zeta_2 < \dotsb$, the $m=1$ modes dominate for $x \gg a$, and
\begin{equation}
  \psi_0(x,0) \sim \alpha_{1} e^{-\zeta_1(x/a-\half)}.
\end{equation}
Comparing this with the asymptotic behavior of our soliton ansatz, $f(x) \sim 2Ae^{-b x}$, we find that $b = \zeta_1/a$.

In the opposite limit, near the center of the X well, $|x| \ll a$, we are to consider the modes in the central region.
A series expansion around $x=0$ gives that $\psi_0 \simeq 2 \left(1 - k^2 x^2/4 \right) \sum_{m,s} \alpha_{ms} (-1)^\frac{m-1}{2} \sech(\half \zeta_m)$.
By comparison to $f(x) \simeq A(1 - (b x)^2/2)$, this means that $k^2/2 = b^2 = (\zeta_1/a)^2$, or
\begin{equation}
  k = \sqrt{\frac{2}{3}}\frac{\pi}{a}.
\end{equation}
This simple estimate of $k$ is, in fact, within $1\%$ of the numerically attained value (at $L=3-30$)!

\begin{figure}
  \includegraphics{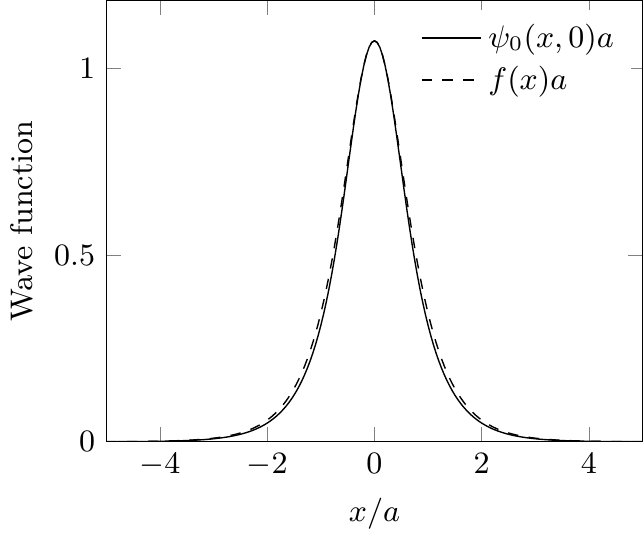}
  \caption{%
    The normalized ground-state wave function $\psi_0(x,0)$ for $L=5$ evaluated along the $x$-axis compared to the soliton approximation $f(x)=A\sech(bx)$ with $b=k/\sqrt{2}$.
    The constant $A$ has been set such that $A=f(0)=\psi_0(0,0)$.}
  \label{fig:gs-soliton-comparison}
\end{figure}

Figure~\ref{fig:gs-soliton-comparison} compares the ground state to $f(x)$ and we see that an extremely good agreement is obtained.
If we try to generalize our soliton ansatz and fit the function $A\sech^p(bx)$ (with fitting parameters $p$ and $b$, and fixed $A=\psi_0(0,0)$) to the numerical wave function of our ground state along the $x$-axis, we arrive at the stable (against starting point guesses) fit $p \simeq 1$, $ba = 1.8-1.9$.
This is in agreement with our original ansatz with $p=1$ and $b=k/\sqrt{2}=1.8050/a$.
From this point forward, we shall often refer to a localized state as a soliton.

\subsection{Variational approach: Excited states}

\begin{figure}
\includegraphics{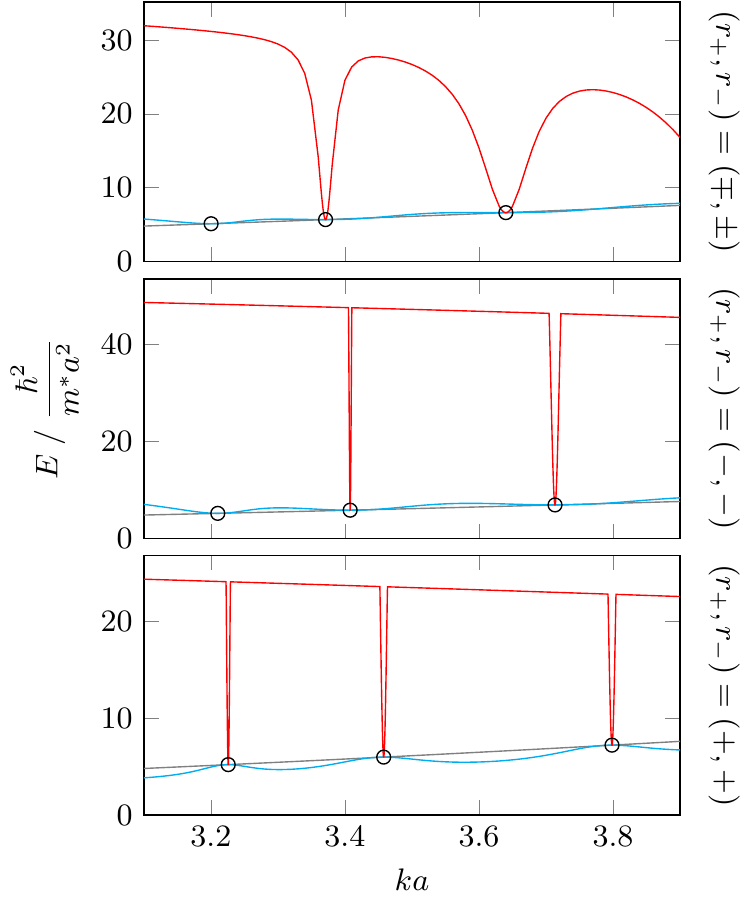}
\caption{%
  Energy of the optimum mode expansion as a function of wave number.
  Each panel correspond to a different symmetry family $(r_+,r_-)$.
  The cyan curve shows the energy of an unconstrained mode expansion.
  The red curve shows the energy of a mode expansion that is orthogonalized to the lowest state in the given symmetry family.
  The grey curve shows the function $E=\hbar^2k^2/2\effm$ for comparison.
  The pairs $(k,E)$ that correspond to actual energy eigenstates are encircled.}
\label{fig:mode-expansion-energy}
\end{figure}

The symmetries ensure orthogonality between states of different family, and it follows that variational bounds are in fact obeyed for the lowest states in each family (despite our previous remark that this cannot be guaranteed in general) provided we enforce the particular symmetry in the mode expansion.

Requiring a given state to be orthogonal to all lower-lying states as prescribed by Section~\ref{sec:xwell-excited-states} often gives rise to very sharp minima in $E(k)$ that can be difficult for a minimization algorithm to locate.
Because of the high level of symmetry, however, it is, in practice, sufficient to find local extrema in the energy of an unconstrained mode expansion with the desired symmetry.
The $(\pm,\mp)$ states and the $(-,-)$ states appear as local minima in the $(k,E)$-curve, whereas the $(+,+)$ states in this energy range appear as local maxima (with the ground state being an exception to this).
See Fig.~\ref{fig:mode-expansion-energy}.
We verify the result of this by comparing to a local minimization with forced orthogonality to the lowest state in the given symmetry family (using the result of the previous calculation as starting point).
With an absolute tolerance of $10^{-6}/a$ in the downhill simplex routine, the two methods are found not to differ by more than $\num{2e-6}/a$ in the resulting wave number $k$.

Notice on Fig.~\ref{fig:mode-expansion-energy} that the energy minima in the orthogonalized mode expansion can be rather sharply located about an eigenstate wave number.
This tendency worsens if the mode expansion is orthogonalized to multiple states.

\begin{figure*}
  \includegraphics{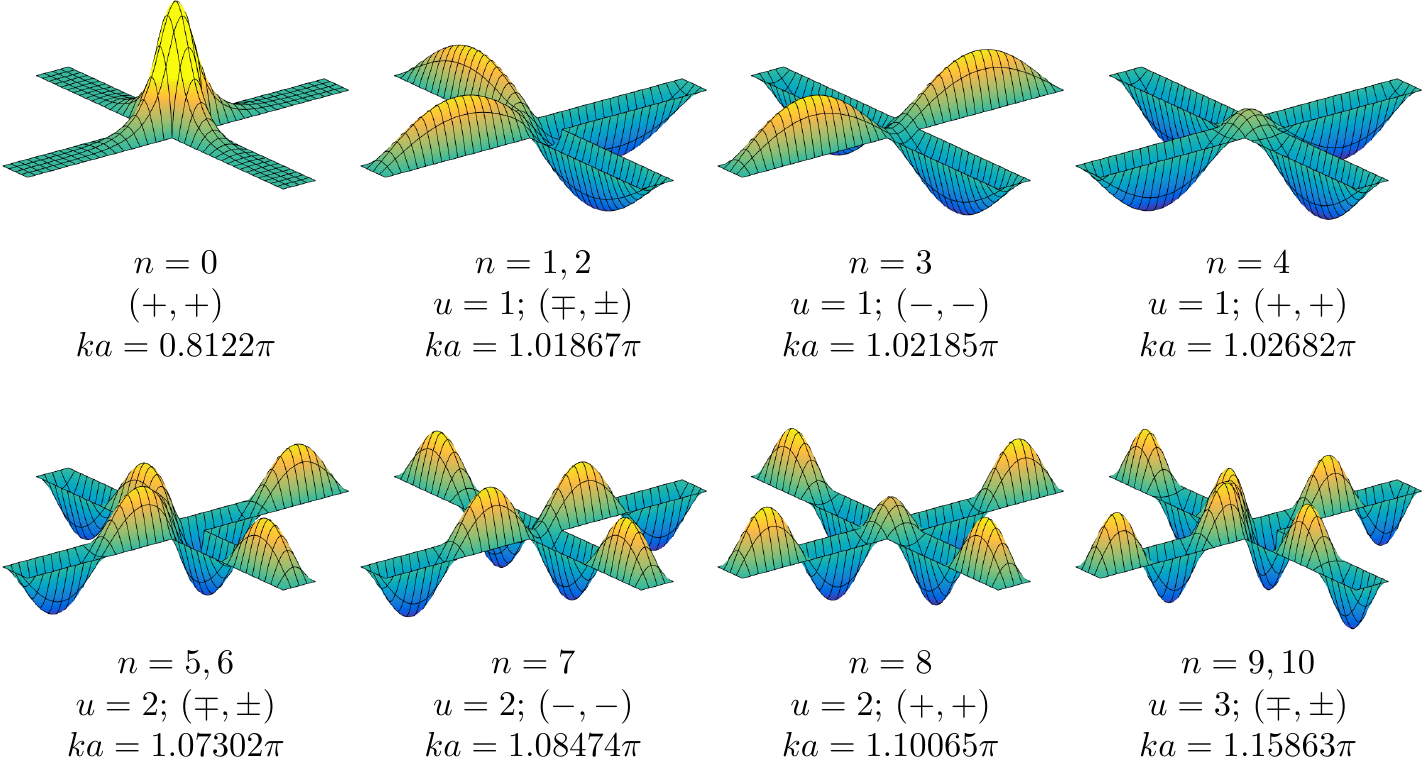}
  \caption{%
    Plots of eigenstate wave functions (vs.\ $x$ and $y$) of the symmetric X well with all legs having length $L_s=5$.
    The plots are labeled with their excitation number $n$, (for the excited states) the $u$ multiplet they belong to (cf.~the discussion preceding Eq.~\eqref{eq:box-energy}), their symmetry family $(r_+,r_-)$ and their wave number $k$.
    A basis size of $M=110$ modes has been used to compute the wave numbers.
    States with symmetry $(-,+)$ and $(+,-)$ are degenerate and their wave functions are identical, save for a rotation of $\pi/2$ in the $xy$-plane.
    The shown plots are consecutive eigenstates from the ground state to the tenth excited state.
  }
  \label{fig:symmetric-xwell-plots}
\end{figure*}

\begin{figure}
\includegraphics{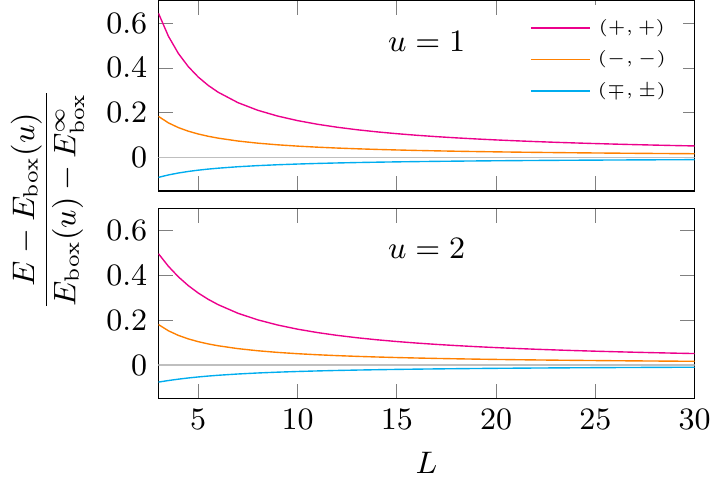}
\caption{%
    The plots show that as $L$ increases, the eigenenergies of the X-well excited states, $E$, approach the energies of an $a \times aL$ box, $E_\text{box}(u)$, faster than the convergence of the latter towards their limit $E_\text{box}^\infty \equiv {\hbar^2\pi^2}/{2\effm a^2}$. %
    The upper panel shows the states converging to $E_\text{box}(1)$, while the lower panel shows those converging to $E_\text{box}(2)$. %
    The legend identifies the eigenstates by the pair $(r_{+},r_{-})$ and applies to both panels.}
\label{fig:excited-state-energy}
\end{figure}

The first few excited states are plotted in Fig.~\ref{fig:symmetric-xwell-plots} for $L=5$.%
\footnote{%
  To avoid cluttering the plots, we have decided at times to omit the axes from surface plots of wave functions.
  Acknowledging that this may be a cause of confusion, the plots have been carefully standardized, as described in the Supplemental Material.
  The surface plots are solely intended for visualising the wave functions; they are not suitable for making accurate readings of $\psi(x,y)$.
  Therefore, surface plots without axes are supplemented with contour plots in the Supplemental Material.%
  \label{fn:no-axis-surface-plots}%
}
With double-precision numbers, we are able to compute expansions up to $M=110$ modes.
This is sufficient for the wave number of the excited states to converge to 5 decimals.
The ground-state wave number is converged to 4 decimals -- the ground state appears to converge more slowly due to it being localized about the central region.
In the following, if not stated otherwise, we shall use mode expansions of $M=30$ modes, which is typically enough for the energies and wave functions to have reasonably converged.

We notice from the transverse part of the wave functions that the first mode ($m=1$) appears to be dominating.
This is so because the available energy is insufficient to appreciably excite $m>1$ modes.
Due to the required orthogonality to the ground state, the $(+,+)$ states inherit a `bump' in their wave function at the center of the X well, as seen in Fig.~\ref{fig:symmetric-xwell-plots}.
This bump, however, diminishes with increasing $L$.

The wave functions resemble the stationary solutions to the two-dimensional particle-in-a-box problem with a box of dimensions $a \times aL$.
We can group the excited states into multiplets of near-degenerate states characterized by the longitudinal excitation number $u$ of the particle-in-a-box state they look like.
The approximate energy of states within a multiplet $u$ is
\begin{equation}
  E_\text{box}(u) = \frac{\hbar^2}{2\effm} \frac{\pi^2}{a^2} \left( 1 + \frac{u^2}{L^2} \right). \label{eq:box-energy}
\end{equation}
Figure \ref{fig:excited-state-energy} confirms that, indeed, $E \simeq E_\text{box}(u)$ for the lowest eight excited states.
The approximation improves with increasing $L$ as the central region of the X well becomes relatively less important.

The energy of states with $r_x=r_y=+1$ is necessarily somewhat larger than $E_\text{box}(u)$ because the wave function approaches its central value from the same side in two opposing legs.
This increases the curvature of the wave function around the X-well center and, thus, the kinetic energy.

For the $E$ representation of the $\dihedral4$ symmetry (see Table~\ref{tbl:D4-char-table}), a mode expansion with only a single mode has a minimum energy of exactly $E_\text{box}(u)$, since we can prepare a solution with that exact energy in the box 
$\{|y|\le\half a\}$ and nothing outside.
This mode expansion has no contribution to the energy by the kinks.
As we increase the number of modes from one to many, we must therefore obtain an energy that is smaller than $E_\text{box}(u)$, since an expansion with more modes has a smaller energy and because the total available area in the X well is larger than $a\times 2aL$.

\subsection{Other localized states}

\begin{figure}
\includegraphics{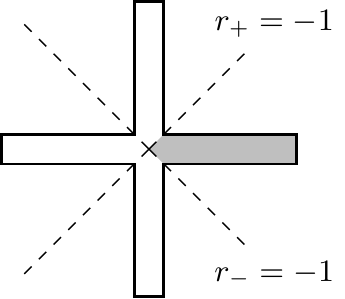}
\caption{%
  For states with $(r_+,r_-)=(-,-)$, the four legs of the X well decouple.
  To find such states, it is sufficient to consider solutions to the Helmholtz equation that vanish outside the shaded pentagon.}
\label{fig:subwells-by-symmetry}
\end{figure}

In this section, we investigate whether the X well supports other localized states than the ground state.
Such localized states would be embedded in the spectrum of excited states. 
We note that for systems with open boundary conditions at the end of the legs, such 
localized states at higher energies may accordingly be classified as bound states in the continuum \cite{hsu2016}.
Our approach may be used also for open boundaries with minute modiciations and could thus study such states as well.

States with symmetry $(r_+,r_-)=(-,-)$, that is, the representations $A_2$ and $B_1$, have $y=\pm x$ as nodal lines.
Finding solutions with this symmetry reduces to solving the Helmholtz equation in the geometry shown in Fig.~\ref{fig:subwells-by-symmetry} with Dirichlet boundary conditions.
This system does not support any localized states, by the following argument.

The energy of the ground state in Fig.~\ref{fig:subwells-by-symmetry} cannot be smaller than the ground-state energy of a system with a larger bounding box.
The geometry is invariant under reflections in the longitudinal axis, so all its eigenstates have either exclusively even or exclusively odd modes $m$.
For odd $m$, the wave function is symmetric under reflection and the threshold for localization is $E_\text{th} = \hbar^2\pi^2/2\effm a^2$.
The smallest rectangle that encloses the pentagon in Fig.~\ref{fig:subwells-by-symmetry} has ground-state energy $E_\text{box}(1)$.
As $E_\text{box}(1) > E_\text{th}$, the solutions to the pentagon cannot be localized unless the lowest mode is completely depleted, i.e., $\alpha_1=0$, by accident.
A similar argument applies to the modes with even $m$, considering only half the pentagon instead.

The representation $B_2$ has $r_x=-1$ and $r_y=-1$, so in this case the problem reduces to finding the eigenstates of an L-shaped well with legs of length $aL$ and width $a/2$.
This system does, in fact, have a localized state whose wave number is $k=1.93\pi/a$ \cite{exner1989,schult1989}.

Numerical analysis reveals that the $E$ representation does not have any localized states in the energy range of interest \cite{delitsyn2012}.

In summary, the X well has one localized excited state, but its energy is so high that it is irrelevant to our purposes of 
studying low-energy dynamics in the next section.

\section{Wave propagation}
\label{sec:xwell-wave-propagation}

\begin{figure}
  \centering
  \includegraphics{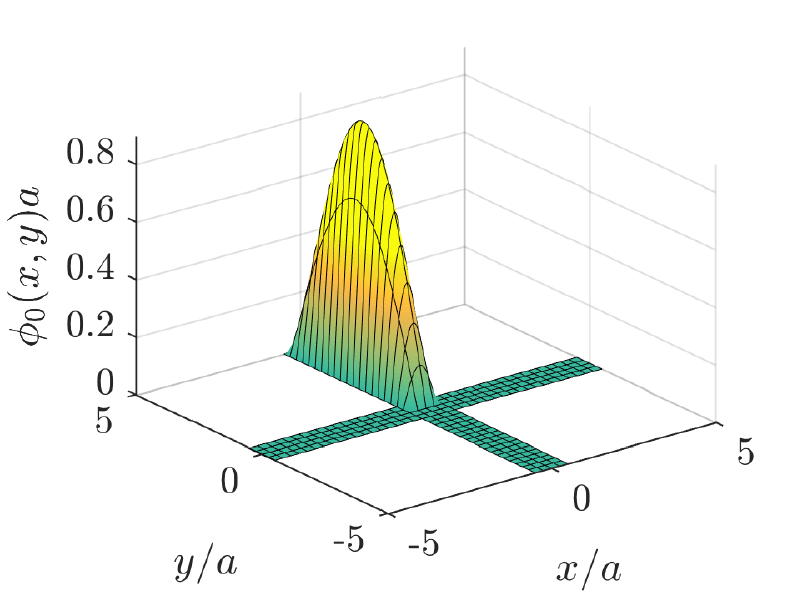}
  \caption{%
    A surface plot of the wave function $\phi_0$ of an initially prepared state in the northern leg.
    In the shown example, $L=5$.
  }
  \label{fig:legwave-initial-state-plot}
\end{figure}

In this section, we investigate how the individual legs of the X well couple and how a wave incident on one leg propagates to the other legs.
This is especially interesting in an application where the X well forms part of a larger network of wires and one wants to send signals of information through the network.
We assume a symmetric X well for simplicity.

Imagine that the crossing of the legs is blocked by some potential that completely decouples the four legs.
We place a particle in one of the legs -- say, the northern leg.
The particle is prepared in its ground state $\ket{\phi;0}$, whose wave function will look approximately like that in Fig.~\ref{fig:legwave-initial-state-plot} with the details depending on the specific blocking potential.
The prepared state in Fig.~\ref{fig:legwave-initial-state-plot} is simply the ground state of an $a\times aL$ box:
\begin{equation}
  \phi_0(x,y) = \braket{x,y|\phi;0} = \frac{2}{a\sqrt{L}} \sin\!\big(\frac{\pi}{aL} y\big) \sin(\pi(\half+x/a))
\end{equation}
for $|x|\le\half a$ and $y\ge 0$, and $\phi_0(x,y)=0$ elsewhere.
The three other legs are initially empty.

Now, at a time $t=0$, the potential barrier in the center is removed instantaneously (relative to the characteristic time-scale of the system) and the particle in the northern leg is free to propagate around the X well.
Technically, this is achieved by expanding the initial state in X-well eigenstates and time evolving the eigenstate expansion.
Thus, 
\begin{equation}
  \ket{\phi;t} = \sum_n e^{-i E_n t/\hbar} \braket{\psi_n|\phi;0} \ket{\psi_n},
\end{equation}
denoting the energy eigenstates $\ket{\psi_n}$ and their energy $E_n$.
The similarity of the time-evolved state to the initial state may be described by the correlation amplitude
\begin{equation}
  C(t) = \braket{\phi;0|\phi;t} = \sum_n e^{-i E_n t/\hbar} \, |\!\braket{\psi_n|\phi;0}\!|^2.
\end{equation}
If the norm of $C(t)$ is close to one, $\ket{\phi;t}$ and $\ket{\phi;0}$ are alike.
If the norm of $C(t)$ is close to zero, on the other hand, $\ket{\phi;t}$ is almost orthogonal to the initial state.

\begin{figure}
\includegraphics{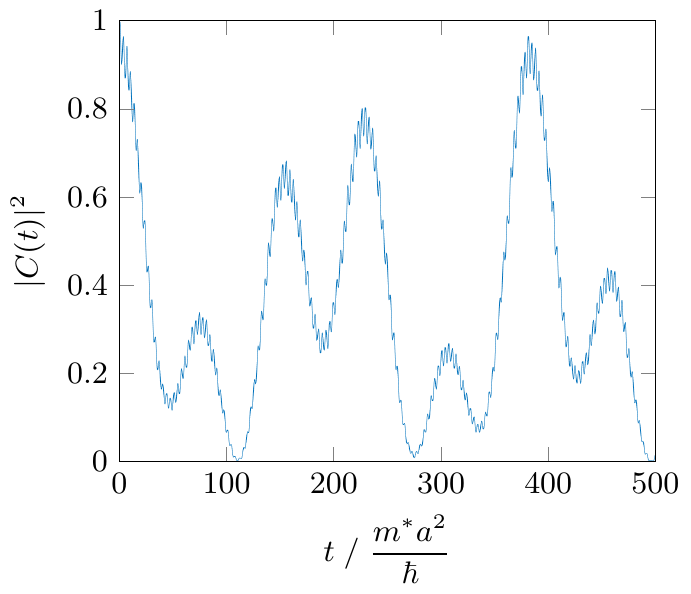}
\caption{%
  The squared norm of the correlation amplitude $C(t)$ plotted against time $t$ for a wave initially confined to a single leg.
  Here, $L=5$ and the initial state has been expanded in the first nine eigenstates, which are in turn found as mode expansions in $M=30$ modes.
  The probability undescribed by the eigenstate expansion, that is, the tail of the sum $\sum_n |\!\braket{\psi_n|\phi;0}\!|^2$, is $\num{0.0013}$, so the initial state is rather well accounted for.}
\label{fig:legwave-correlation-amplitude}
\end{figure}

The correlation amplitude is plotted in Fig.~\ref{fig:legwave-correlation-amplitude} as a function of $t$.
We see from the figure that $|C(t)|^2$ oscillates.
The oscillations contain both high and low frequencies.
The high-frequency oscillations have a period $T_\text{high}\simeq3.4\,{\effm a^2}/{\hbar}$ and are primarily due to the overlap with the ground state.
Indeed, $\ket{\phi;t}$ has mean energy
\begin{equation}
  \braket{E} = \sum_n E_n \,|\!\braket{\psi_n|\phi;0}\!|^2 = 5.1098\,\frac{\hbar^2}{\effm a^2},
\end{equation}
from which the ground-state oscillation period is found to be $2\pi\hbar/(\braket{E}-E_\text{gs})=3.39\,{\effm a^2}/{\hbar}$.

The low-frequency oscillations with period $T_\text{low}\simeq77\,{\effm a^2}/{\hbar}$ are mostly due to the overlap with the first four excited states.

At some points in time ($t=100, 275, 496\,{\effm a^2}/{\hbar}$ in the region plotted), the correlation amplitude is close to zero.
This is because the wave $\ket{\phi;t}$ has propagated onto the southern leg and has almost no overlap with the initial state.
At the two small tops about $t=75\,{\effm a^2}/{\hbar}$ and $t=307\,{\effm a^2}/{\hbar}$, $|C(t)|^2\simeq\SI{25}{\%}$, meaning that all four legs are more or less equally populated.

Finally, at $t=381\,{\effm a^2}/{\hbar}$, we have an almost complete revival of the original state (lacking only $|C(0)|^2-|C(t)|^2=\num{0.0319}$).

\begin{figure}
\includegraphics{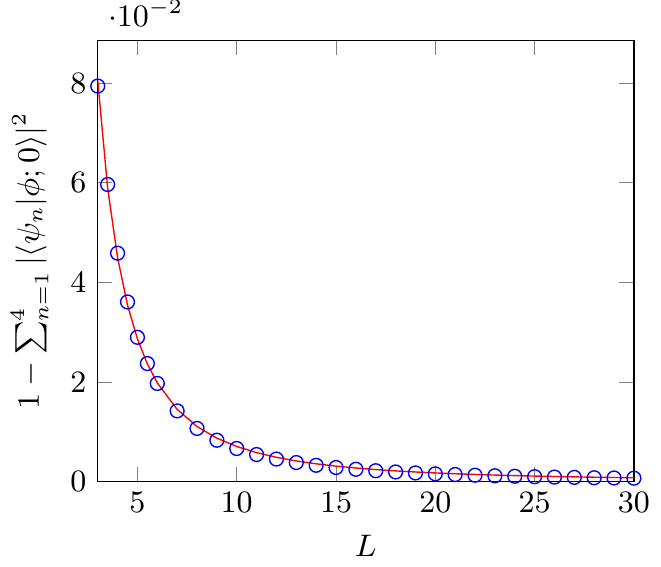}
\caption{%
  Missing probability in the expansion of the initial wave $\ket{\phi;0}$ solely in the first four excited states.
  Notice the scale of the ordinate axis.
  The solid curve is a fit to a power law.
  The fit is consistent with an exponent of $-2$.}
\label{fig:legwave-missing-probability}
\end{figure}

The initial state $\ket{\phi;0}$ is described almost exclusively by the four lowest excited states.
This statement becomes closer to being true as the size of the X well is increased and with it the relative importance of the central region decreased.
From Fig.~\ref{fig:legwave-missing-probability}, it appears that $\sum_{n=1}^4 |\!\braket{\psi_n|\phi;0}\!|^2 \to 1$ as $L \to\infty$.
A fit to a power function further shows that
\begin{equation}
  1 - \sum_{n=1}^4 |\!\braket{\psi_n|\phi;0}\!|^2 \sim \frac{1}{L^2}. \label{eq:legwave-expansion-convergence}
\end{equation}

The ground state is independent of $L$ and is primarily confined to a region $\{|x|,|y|\le a R\}$, where $\half<R\ll L$.%
\footnote{It is not essential what $R$ is, but we could take it to be, e.g., $R=\max\{\sqrt{x^2+y^2} \vert \braket{x,y|\psi_0} \ge 0.1\braket{0,0|\psi_0}\}$.}
The inner product between the initial state and the ground state scales approximately as the overlap between the initial state and a constant function on the region $\{|x|,|y|\le aR\}$:
\begin{multline}
  \braket{\psi_0|\phi;0} \sim \int_{-\half a}^{\half a} \dif{x} \int_0^{a R} \dif{y}\; \phi_0(x,y) \\
  = \frac{8}{\pi^2}\sqrt{L}\sin^2\!\bigg(\frac{R \pi}{2L}\bigg) \sim \frac{1}{L^{3/2}}.
\end{multline}
Since, therefore, $|\!\braket{\psi_0|\phi;0}\!|^2 \sim 1/{L^3}$, it is not the ground state, but the higher excited states that limit the convergence, cf.~Eq.~\eqref{eq:legwave-expansion-convergence}.

Neither the ground state nor the higher excited states play any significant role in the long-term time evolution -- recall from Fig.~\ref{fig:legwave-correlation-amplitude} that the contribution from the ground state is small (due to the small overlap) and merely results in rapid oscillations that die out on average.
In particular, the ground state is not responsible for the couplings of the legs.

The small energy differences among the $n=1,\dotsc,4$ states give rise to a beat phenomenon that propagates the wave to the other legs.
For $L=30$, the revival time is $T\sim 10^4\, \effm a^2/\hbar$, giving $T\sim\SI{e-10}{s}=\SI{e-1}{ns}$ if $a=\SI{1}{nm}$ and $\effm$ is the electron mass.
This revival time is large enough that it should be measurable.
The applicability of the X-well model that we have developed in this chapter is, thus, experimentally testable.

\section{Quantum graphs}
\label{sec:quantum-graph-theory}

For an X well whose legs are long and thin, it is typically assumed that only the $m=1$ mode is present due to the high energy requirements for exciting the transverse motion.
As $L\to\infty$ with $aL$ held fixed, the system is claimed to be effectively described as a so-called quantum graph of one-dimensional edges meeting at a central vertex.
The edges are the legs of the X well and the vertex is its center.
If the quantum-graph description is valid, it is advantageous due to its simplicity -- the transverse degree of freedom has been integrated out, so the wave function on the graph depends only on a single parameter.

The dynamics of the quantum graph are governed by the one-dimensional Schrödinger equation of a free particle.
In natural units ($\hbar=\effm=aL=1$), this is
\begin{equation}
  E\, \eta = -\frac{1}{2} \od[2]{\eta}{z_s},
\end{equation}
where $z_s\in[0,1]$ is the distance along the edge $s$ from the vertex at $z_s=0$, and $\eta(z_s)$ is the one-dimensional wave function on $s$.

\begin{figure}
  \includegraphics{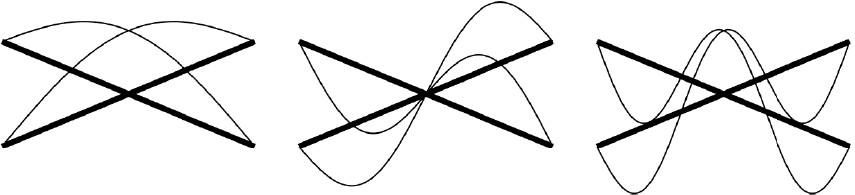}
  \caption{%
    Sketches of three quantum-graph solutions with Kirchoff boundary conditions at the center vertex and Dirichlet conditions at the end points.
    The states have $k=\half\pi$, $\pi$ and $\tfrac32\pi$, respectively.
    The thick line represent the edges of the graph and the thin curves show the wave function $\eta$.}
  \label{fig:star-graph-plots}
\end{figure}

In a seminal paper, \textcite{ruedenberg1953} have shown that conservation of probability current imposes so-called Kirchoff boundary conditions at the vertex \cite{ruedenberg1953,kuchment2008}:
$\eta(z_s)$ is continuous at the vertex and 
\begin{equation}
  \sum_{s} \od{\eta}{z_s}\bigg|_{z_s=0} = 0.
\end{equation}
The sum runs over all four edges that meet at the vertex.
The solutions obeying these boundary conditions are of the form $\eta(z_s) = \sin(k(1-z_s))$ with wave number
$k = \tfrac{1}{2}\pi, \pi, \tfrac{3}{2}\pi, 2\pi, \dotsc$
Examples are shown in Fig.~\ref{fig:star-graph-plots}.

The $k = \half\pi, \tfrac32\pi, \tfrac52\pi, \dotsc$ states predicted by quantum graph theory belong to the $(+,+)$ symmetry family, while the states with $k=\pi,2\pi,3\pi,\dotsc$ belong to the other families since their wave functions have a node at the vertex.

The effective one-dimensional Schrödinger equation together with the Kirchoff boundary conditions, however, do not allow for localized bound states such as the ground state of our X well.
In fact, the derivation of Ruedenberg and Scherr did not take localized states into account \cite{kuchment2002}.%
\footnote{Alternative boundary conditions have been suggested in the literature \cite{voo2006,gratus1994}, but these suggestions do not seem to agree with our results either.}
The existence of the localized ground state forbids the $(+,+)$ states predicted by quantum graph theory (with Kirchoff boundary conditions) as they are not orthogonal to the ground state.
Hence, among the solutions plotted in Fig.~\ref{fig:star-graph-plots}, only the middle one is a true limiting state of the X well.

For the excited states of the X well, we know from our numerical analysis that $E \to E_\text{box}(u)$ as $L\to\infty$.
The longitudinal wave number of the lowest mode becomes $k_{1\parallel}aL\simeq\pi,2\pi,3\pi,\dotsc$
So when going from the two-dimensional description of the X well to the limiting quantum-graph model, all excited states -- including the $(+,+)$ family -- have $k=\pi,2\pi,3\pi,\dotsc$

As $L$ becomes very large, the legs of the X well decouple.
Each multiplet of excited states becomes four-fold degenerate and its members may be linearly combined to form particle-in-a-box states, each only residing in a single leg.
This suggests that the correct boundary condition for an equivalent quantum graph model is to enforce a node at the vertex, that is, $\eta(z_s=0) = 0$ for all edges $s$, but make no restrictions on the derivative of the wave function.%
\footnote{We remark that these observations are in concurrence with the conjecture that the graph decouples in the vicinity of the energy threshold as mentioned in \textcite{cacciapuoti2007}.}
This, of course, still cannot describe localized states such as the ground state.


The ground state is difficult to treat in a quantum graph model, because in the one-dimensional limit $a\to0$, the ground-state density is everywhere zero save for exactly at the vertex.
The wave function may, thus, somewhat come to resemble a Dirac delta function.
In the mathematics literature, this problem is typically circumvented by rescaling the system and considering the analogous limit where $a$ is held constant while $L\to\infty$ \cite{cacciapuoti2007,grieser2008}.

We conclude that the $a\to0$ limit of the X well is rather pathological; the excited states decouple and the ground state is ill-defined.
The dynamics are, thus, trivial.
A real physical system is never truly one-dimensional and will possess dynamics, so quantum graph theory is not necessarily beneficial in obtaining a description of the dynamics of an X well and in turn of larger quantum networks.

\subsection{An effective non-linear Schrödinger equation}

In spite of the fact that localized bound states cannot be described by the linear, one-dimensional Schrödinger equation, such states do exist in quantum graphs governed by the \emph{non-linear} Schrödinger equation \cite{adami2012}
\begin{equation}
  -\frac{1}{2}\od[2]{\eta}{z_s} + U_0|\eta|^2 \eta = \mu \eta
\end{equation}
with $U_0,\mu<0$. These have been proposed for the description of graph geometries occupied by bosons in 
the condensed state \cite{adami2012,markowsky2014}. We find that 
these soliton-like solutions are, in fact, of the form $f(z_s)=A\sech(b z_s)$ that we have shown the ground state approximately to follow, cf.~Section~\ref{sec:soliton-approximation},
provided that $E=-2\mu$ and $A^2=2\mu/U_0$.
It, thus, appears that a single particle trapped in an X well can in some respects behave like a Bose-Einstein condensate.
In terms of cold atoms, it is now possible to realize box potentials for condensates \cite{gaunt2013} and thus if 
one can find a way to cross several such boxes, one may built a potential similar to the X-well or variations of it.
Solitonic excitations are of course known also for other fields such as optical fibers and photonic crystals 
\cite{kivshar1989,drummond1993,kartashov2011} and our developments and formalism could possibly be applied to 
such systems with some modifications that account for the propagation of photons instead of massive particles.

\section{Variations on the X well}
\label{sec:xwell-variations}

In the previous sections, we have restricted our attention to the symmetric X well due to its simplicity of analysis.
However, we stress that the method presented in Section~\ref{sec:xwell:eigenstates-as-mode-expansions} applies generally and do not depend on symmetries of the well.
Also, numerical experiments show that the ground state survives in the general case.
Indeed, the exponential decay of its legs hinders it from `seeing' the end walls of the legs.
Neither is any significant change in the ground state induced if we change the boundary conditions on some of the end walls, e.g., to periodic or open boundary conditions.
The latter case of open boundaries has been studied in some detail in the literature \cite{schult1989,avishai1991,markowsky2014,amore2012}.

\begin{figure}
\includegraphics{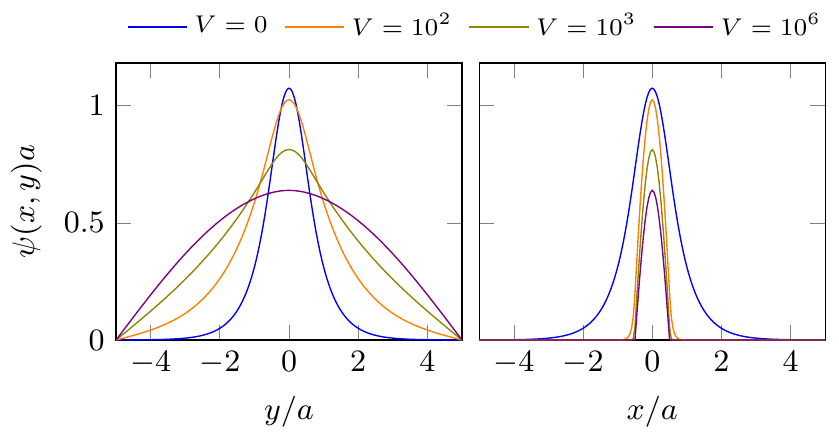}
\caption{%
  Transition from a localized to a non-localized state as the potential offset $V$ of the eastern and western legs is increased.
  The left panel shows the wave function $\psi$ vs.\ $y$ at $x=0$.
  The right panel shows $\psi$ vs.\ $x$ at $y=0$.
  For $V=0$, the X well is symmetrical.
  All four legs are of length $L=5$.
  The legend reports the potential offset in units of $\hbar^2/\effm a^2$ and applies to both panels.}
\label{fig:localization-transition}
\end{figure}

\subsection{Imposing a potential upon a leg}
\label{sec:xwell-potential-offset}

Let us consider how the variational method for finding eigenstates generalizes if an external potential offset is imposed on part of the X well.
This is relevant as such a potential offset may be used to manipulate the particle in the well.

If we impose a constant potential offset $V_s$ upon the leg $s$, along that leg the wave number $k_{s}$ must satisfy $\hbar^2k_{s}^2/2\effm + V_s = E = \hbar^2k^2/2\effm$, meaning that the longitudinal component of the $m$'th mode wave vector is
\begin{equation}
  k_{ms\parallel} = \sqrt{k^2 - \frac{2\effm}{\hbar^2}V_s - \left(\frac{m\pi}{a}\right)^2}
\end{equation}
along the leg $s$.
In the central region, the expression for the wave vector is unchanged.

If $E$ is the energy of an eigenstate in the case $V_s=0$ for all $s$, and we perturb the well by imposing a potential $V$ on one of the legs, we must expect the eigenstate of the perturbed system to have an energy $E'$ in the range between $E$ and $E + V$.
This follows by noting that if we had imposed $V$ on the entire X well, the eigenstates would remain unchanged, only everywhere replacing $k^2$ with $k^2+2\effm V/\hbar^2$.
For small $V$, this bracketing of $E'$ can help in numerically determining the correct wave numbers of the eigenstates of the perturbed system.

Imagine we have an X well with $L_s=L$ for all $s=\rleg,\uleg,\lleg,\dleg$.
If we impose an infinite potential offset on two opposing legs, $V_\rleg=V_\lleg=V=\infty$, then we effectively have $L_\rleg=L_\lleg=0$, and the well is just an $a \times 2aL$ rectangle, whose ground-state wave number is $k=\pi\sqrt{1+1/(2L)^2}/a$.

If we tune the potential offset $V$ of the eastern and western legs from zero to infinity, we must have some crossover from the localized ground state of the symmetric X well to the single-mode particle-in-a-box ground state.
As we see from Fig.~\ref{fig:localization-transition}, the crossover turns out to be continuous.
This suggests that in the limit $L\to\infty$, the localized state is present for any finite value of $V$.

\subsection{Alternate geometry: The T well}

By imposing an infinite potential offset on one of the legs -- say, $V_\dleg=\infty$ -- we effectively remove that leg from the X well and we are left with a T-shaped well.
In the following, we analyse the T well as an example of a generalized X well with a different geometry.
As we shall see, the variational method is still applicable and the qualitative results do not differ much from the symmetric X well.
We further remark that a T well has an application as a constituent in the boundary of a grid of X wells.

If $L_\rleg=L_\lleg$, the T well has a reflection symmetry in the $y$-axis.
As with the symmetric X well, the symmetry is useful in classifying the energy eigenstates and in the numerical procedures used to compute them.

\begin{figure*}
  \includegraphics{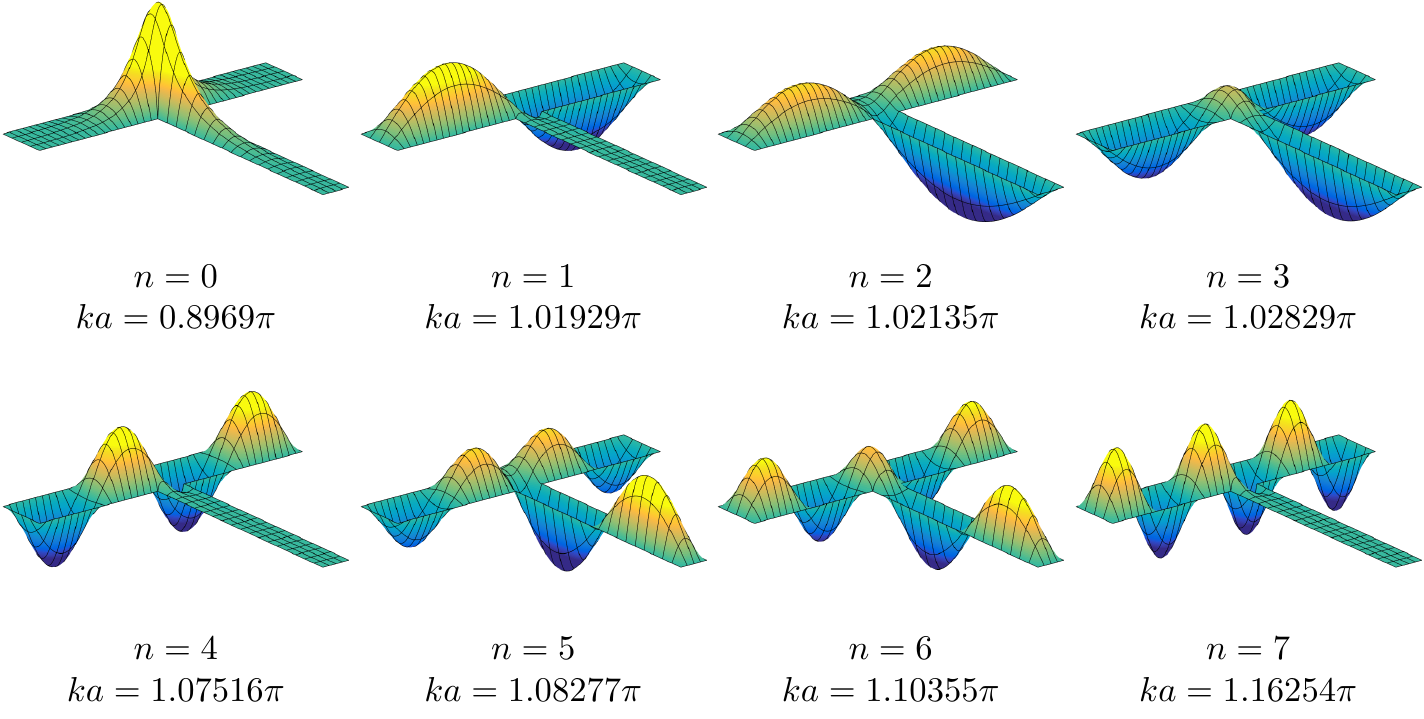}
  \caption{%
    Surface plots of the ground state and first few excited states in a T well with $L_\rleg=L_\lleg=L_\uleg=5$.
    The states are labeled by excitation number $n$ and wave number $k$.
  }
  \label{fig:twell-plots}
\end{figure*}

With a variational method similar to the one employed for the X well, we arrive at the eigenstates plotted in Fig.~\ref{fig:twell-plots}. 
We remark, in particular, that the T well also supports a localized ground state.
We also see that though we have broken almost all of the symmetry of the X well in removing the southern leg, the eigenstate wave functions show many of the same features as in the symmetric X well.

The eigenstates arrange themselves into multiplets of three states whose energies lie close to one another.
The states $n=1,2,3$ constitute one multiplet, $n=4,5,6$ are another multiplet etc.
The three states in a multiplet belong to three different families of states analogous to the $(r_+,r_-)$ families used in the classification of the eigenstates of the symmetric X well.
The analogy appears because, for all of the plotted excited states, the wave function $\psi(x,y)$ in the region $\{|y|\le\half a\}$ is almost symmetric under reflection in the $x$-axis.
Thus, if we re-attached the southern leg and let $\psi(x,y)=\psi(x,-y)$ for $y<\half a$, we would approximately obtain the X-well eigenstates.

The T well has no rotational symmetry, so the two degenerate X-well states $(-,+)$ and $(+,-)$ are reduced to one state in the T well.
In other words, the symmetry group of the T well is Abelian and its irreducible representations are, thus, all one-dimensional.

\subsection{Alternate boundary conditions: The looped X well}

\begin{figure}
  \includegraphics{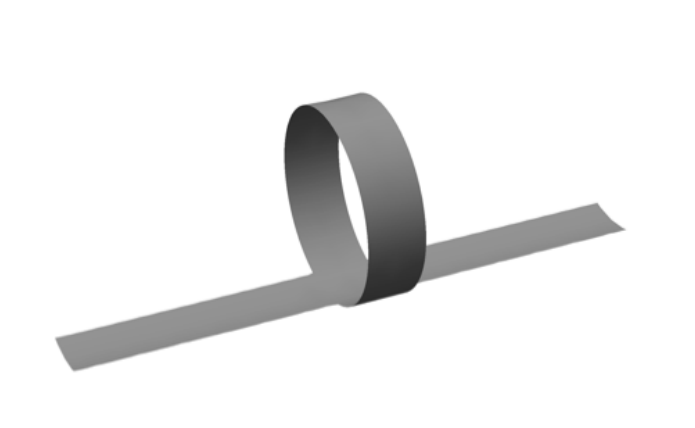}
  \caption{Three-dimensional sketch of a looped X well with $L_\rleg=L_\lleg=5$ and $L_{\uleg\dleg}=10.5$.}%
  \label{fig:looped-xwell-3d-image}%
\end{figure}

\begin{figure}
  \includegraphics{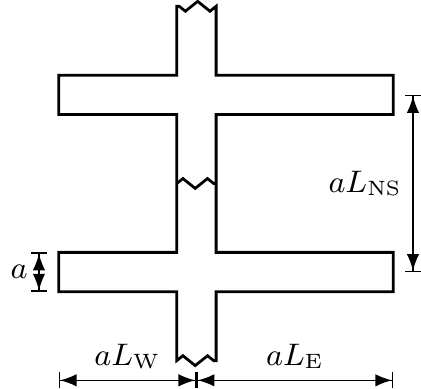}
  \caption{Schematic of a looped X well. The zig-zag lines indicate that the pattern repeats itself.}%
  \label{fig:looped-xwell-schematic}%
\end{figure}

As an example of a generalized X well with different boundary conditions, 
take two opposing legs of the X well -- say the northern and southern legs -- and weld them together.
We arrive at the configuration shown in Fig.~\ref{fig:looped-xwell-3d-image}.
The ring has circumference $L_{\uleg\dleg}=L_\uleg+L_\dleg$.
Looking at the figure, one realizes that this geometry could be sensitive to magnetic fluxes
threading the loop, and potentially be used in sensing of magnetic fields. 

Technically, the join of the two legs is achieved by changing the boundary conditions on the end walls of the legs from closed (i.e., Dirichlet conditions) to periodic.
An outline of the well is shown in Fig.~\ref{fig:looped-xwell-schematic}.

The $\ket{m,\uleg}$ and $\ket{m,\dleg}$ modes are now placed on top of one another, meaning that they may have a non-zero wave-function overlap from the leg-part of the ring.
The matrix elements of $\Psi$ and $\Pi$ between $\ket{m,\uleg}$ and $\ket{m,\dleg}$ modes are different from those of the flat (i.e., non-looped) X well.

We generally expect the eigenstates of the looped X well to have lower energy than for the flat X well since the end constraints on the northern and southern legs are lifted.

For relatively large $L_{\uleg}$ and $L_{\dleg}$, the ground state of the flat X well does not `see' the ends of the legs, so it does not matter whether we join them together.
Therefore, the ground states of the flat and the looped X well are the same.
Contrary to this, one might expect that the ground-state wave function of the looped X well should be constant along the ring.
Such a state has no kinetic energy in the longitudinal direction along the ring, so its energy is exactly the energy contribution due to the lowest transverse mode, $E=\hbar^2\pi^2/2\effm a^2$.
However, as we know, this energy is larger than that of the localized X-well ground state.
Localization is a two-dimensional phenomenon with no obvious analogue in one dimension.

Numerical experiments show that if the length of one of the legs $L_{\uleg\dleg}$, $L_\rleg$ or $L_\lleg$ is less than $\sim 5$, the ground state begins to feel the ends of the legs and the energy becomes length-dependent.
The energy increases when $L_\rleg$ is reduced, but remarkably, it falls when the circumference $L_{\uleg\dleg}$ is lowered.

\begin{figure*}
  \includegraphics{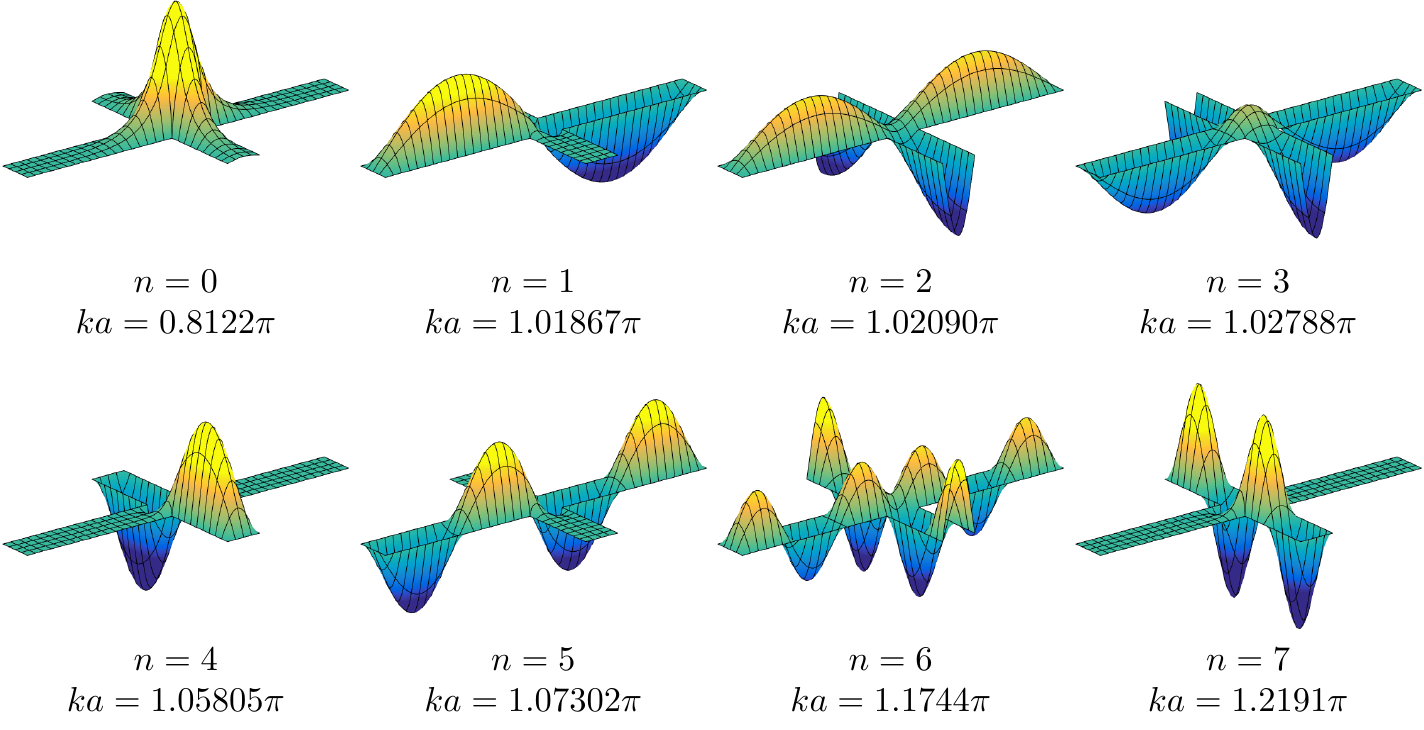}
  \caption{%
    Surface plots of eigenstates in the looped X well plotted for $|y|\le a L_{\uleg\dleg}/2$. The well has $L_{\uleg\dleg}=5.5$ and $L_\rleg=L_\lleg=5$.
    The states are labeled by excitation number $n$ and wave number $k$.
  }
  \label{fig:looped-xwell-plots}
\end{figure*}
Figure~\ref{fig:looped-xwell-plots} plots the eigenstates of a looped X well. 
The wave functions are normalized in the plotted area.

Due to the symmetry, for an eigenstate wave function $\psi$ either
\begin{equation}
  \psi\big(x,\pm a\frac{L_{\uleg\dleg}}{2}\big)=0 \quad\text{or}\quad
  \pd{\psi}{y}\bigg|_{y=\pm a\frac{L_{\uleg\dleg}}{2}} = 0 \label{eq:looped-xwell-boundary-relations}
\end{equation}
must hold (and we remember that the wave function must at any case have the same value and derivative at $y=\pm aL_{\uleg\dleg}/2$).
Furthermore, the states possess reflection symmetry in both the $x$- and the $y$-axis.

As for the T well, the first three excited states resemble the eigenstates we know from the flat X well.
From there on, it gets a little more complicated as states with the two boundary relations Eq.~\eqref{eq:looped-xwell-boundary-relations} mix in among each other.
States for which the first equality in Eq.~\eqref{eq:looped-xwell-boundary-relations} hold are a subset of the solutions to an X well whose northern and southern legs have length $L_{\uleg\dleg}/2$.


\begin{figure}
\includegraphics{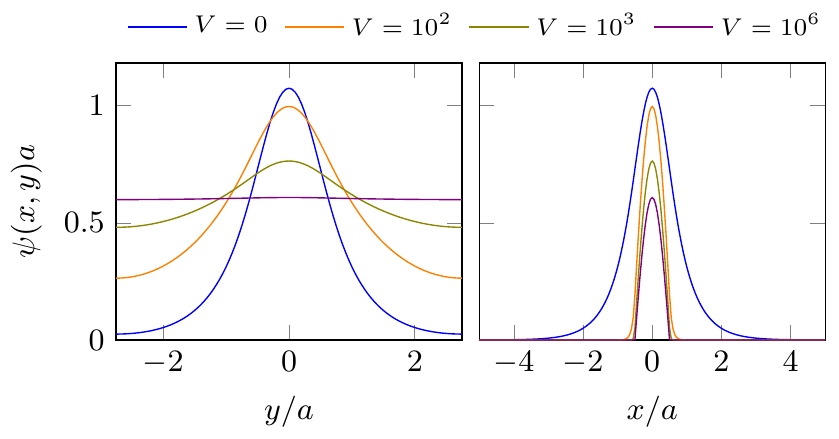}
\caption{%
  Transition in a looped X well ($L_{\uleg\dleg}=5.5$, $L_\rleg=L_\lleg=5$) from the localized X-well ground state to a state that is confined to the ring and constant along it.
  The panels show a cross section of the wave function $\psi$ at $x=0$ and $y=0$, respectively, for different values of the potential offset $V$ of the eastern and western legs.
  The legend reports $V$ in units of $\hbar^2/\effm a^2$ and applies to both panels.
  See also Fig.~\ref{fig:localization-transition} for the analogous experiment in a flat X well.}
\label{fig:localization-transition-looped}
\end{figure}

In order to further check the looped setup, we consider how the system can transition to a ring-confined geometry and how its ground state changes accordingly. Imagine we impose a tunable potential offset on the eastern and western legs, $V=V_\rleg=V_\lleg$.
In Fig.~\ref{fig:localization-transition-looped}, we show that by adiabatically increasing this offset, the ground state transforms continuously into a state that is constant along the ring in the longitudinal direction, has the shape of a sine in the transverse direction and is zero outside the ring.
This is the ground state of a naked ring without any legs extruding from it.

\section{A network of X wells}
\label{sec:xwell-network}

By joining several X wells together, we can create a grid of wires.
The boundaries can be filled with T wells and the corners with L wells.
We are hereby in a position to describe an entire grid.
We have already considered how a wave might propagte through the network, but what are the dynamics of a soliton in the grid?
To answer this question, we consider the simplest possible network of X wells, namely, a double X well.

\begin{figure}
\includegraphics{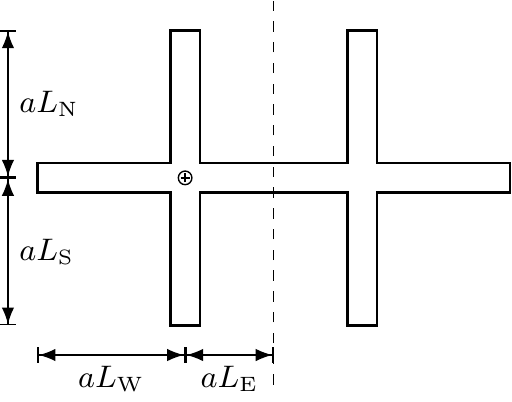}
\caption{%
  Schematic of a double X well with mirror axis (dashed).
  The origin of the $(x,y)$ coordinate system is taken to be at the center of the left X well as indicated by the crosshair on the figure.}
\label{fig:double-xwell-schematic}
\end{figure}

Place two X wells next to each other such that one leg of either well are joined together at the ends, cf.~Fig.~\ref{fig:double-xwell-schematic}.
Assume that the two-well setup is symmetric under reflection in the line $x=a L_\rleg$ halfway through their common leg.
(The important case of joining two \emph{symmetric} X wells is covered by this assumption.)

The reflection symmetry in $x=a L_\rleg$ implies that 
\begin{multline}
  \psi(a L_\rleg,y)=0 \quad\text{(Dirichlet)}\\\text{or}\quad
  \pd{\psi}{x}\bigg|_{{x=a L_\rleg}} = 0 \quad\text{(Neumann)}\label{eq:double-xwell-boundary-relations}
\end{multline}
for odd and even states, respectively.
By symmetry, we only have to consider the left half of the well if we make sure to impose Eq.~\eqref{eq:double-xwell-boundary-relations} as a boundary condition on the eastern end wall.
The odd states correspond exactly to the eigenstates of a single, isolated X well.
This leaves only the even states to be determined.
The Neumann boundary condition on the eastern leg is satisfied with the mode wave function
\begin{multline}
  \braket{x,y|m,\rleg} = \sec(\kpar a(L_\rleg-\half)) \\
  \cdot \cos(\kpar(aL_\rleg-x)) \sin(\kper(\half a+y))
\end{multline}
along the eastern leg.

The ground state of the double X well is expected to be even under reflection.
As we know that the single-well ground state gives rise to an odd, localized state, the double-well ground state must also be localized if it is to have a lower energy than the former.

\begin{figure*}
  \includegraphics{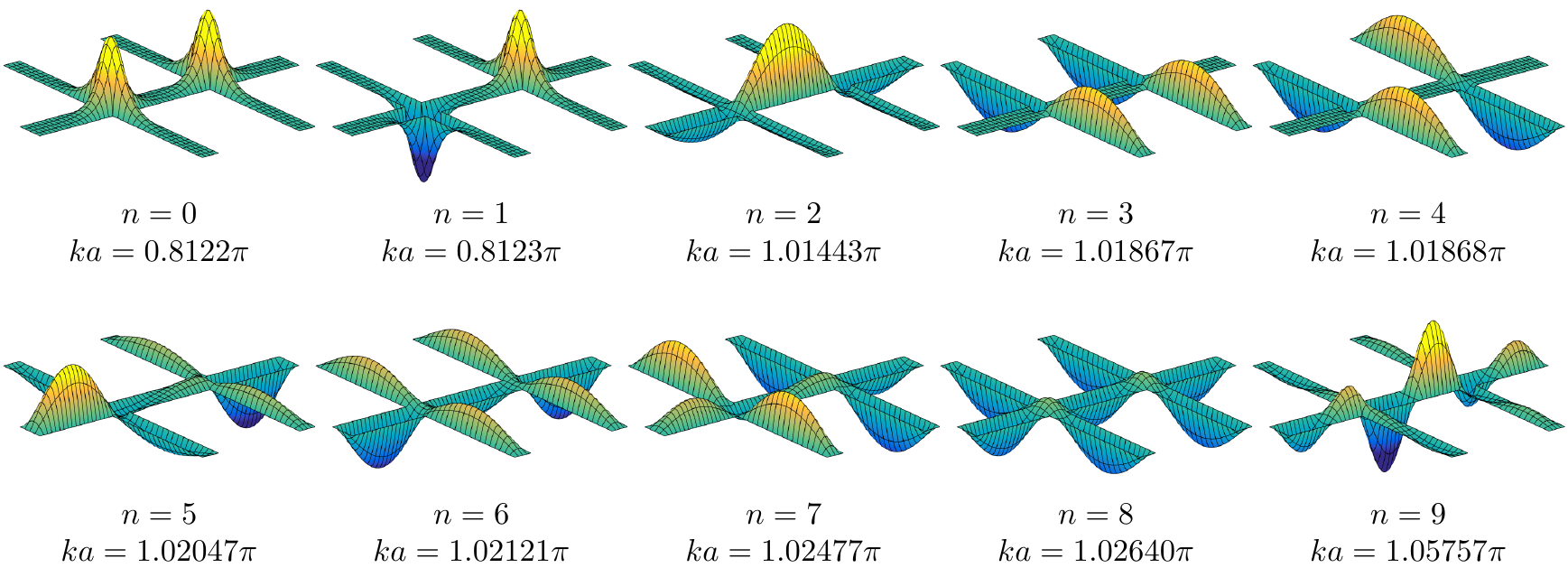}
  \caption{%
    Surface plots of double-X-well eigenstates ($L_\uleg=L_\lleg=L_\dleg=5, L_\rleg=3$).
    The states are labeled by their excitation number $n$ and their wave number $k$.
  }
  \label{fig:double-xwell-plots}
\end{figure*}

Figure~\ref{fig:double-xwell-plots} shows the lowest eigenstates of the double X well found by variation.
The two lowest eigenstates are very much alike and their energies are close, but while the wave function of the even state is  exponentially suppressed at $x=a L_\rleg$, the odd state vanishes exactly.
The energy difference is determined by the wave-function overlap between two solitons (i.e., single-well ground states) prepared in their respective vertex.
The overlap falls exponentially with the distance between the vertices.
If the distance is long enough, the eigenstates are degenerate and decouple into two single-well ground states. 

Extending this observation from the two connected X wells to a large network of X wells, we see that if the distance between neighboring sites is large and the system has been cooled below the excitation threshold ($E_\text{th}={\hbar^2\pi^2}/{2\effm a^2}$), we have realized a lattice.
The simulations of the following section show that a soliton prepared in a site does not couple to the excited states above threshold but only to solitonic states in neighboring sites.
This means that we can describe the system as a discrete lattice with some amplitude for a soliton to hop from one site to another, as depicted in Fig.~\ref{fig:xwell-schematic}(c).

\subsection{Inter-well propagation of a soliton}
\label{sec:inter-well-soliton-propagation}

Prepare a soliton in the left X of a double X well at a time $t=0$ and allow it to propagate under time evolution.
The soliton has a large overlap with the two lowest eigenstates in the spectrum of the double X well, but practically no overlap ($<\num{e-8}$) with the non-localized excited states.
This means that as time evolves, probability density gradually disappears from the left vertex and simultaneously reappears at the right vertex until the soliton has been completely transferred to the right X well.

\begin{figure}
\includegraphics{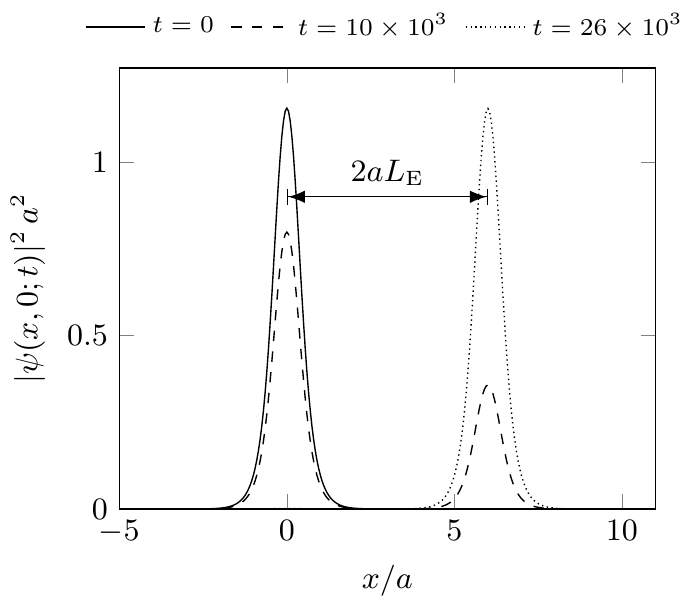}
\caption{%
  Snapshots of the probability density $|\psi|^2$ of a soliton at $y=0$ as it propagates from one well to another.
  The legend reports the time $t$ in units of $\effm a^2/\hbar$.
  The soliton is initially prepared in the left well at $t=0$.
  The double X well has parameters $L_\uleg=L_\lleg=L_\dleg=5$, $L_\rleg=3$.}
\label{fig:soliton-in-double-xwell-cross-section}
\end{figure}

A cross section of the probability density is plotted in Fig.~\ref{fig:soliton-in-double-xwell-cross-section} at three different times during the transfer process; at the beginning, at an intermediate point and at the end of the transfer.
The energy difference between the two localized states of the double X well is very small (cf.~Fig.~\ref{fig:double-xwell-plots}), so the transfer process takes a long time on the natural timescale of the system.
The fact that the soliton never couples to the non-localized states means that barely any probability density is ever found in the 
(outer parts of the) legs.
Once the transfer process has completed, if the system is left to itself, it will begin the reverse process, ending with a revival of the initial state, after which the whole process repeats itself.

The tunneling of these localized waves from site to site is reminiscent of electrons that are tightly bound to ions in a solid where
tunneling happens through the barriers of the potential landscape created by the ions. In the realm of cold atoms, it reminds 
us of the insulator states with exponentially suppressed hopping of atoms between different sites in an optical lattice.

\section{Conclusion and outlook}

We have developed a method of mode expansions to effectively compute eigenstates of the X well with its numerous variations (the T well, looped X well, double X well etc.).
The method has proven very general and is applicable to other two-dimensional geometries constituted by unions of rectangular regions.
We have mainly used a variational method, but have also shown that an approach that explicitly forces a continuous derivative of the wave function may be employed.

We have found that the ground state is localized about the center of the X for all the systems we have considered.
While the existence of a localized ground state is known in the case of open boundary conditions, it has not previously been demonstrated that its wave function is practically independent of the boundary conditions of the end walls.

We have further shown that a cross section of the ground-state wave function along two opposing legs has the same form as a solitonic solution to the non-linear Schrödinger equation.
This enables us to predict the wave number of the ground state to be $k \simeq \sqrt{{2}/{3}}\,{\pi}/{a}$.
When combining several X wells, each vertex supports a localized soliton-like state and these states couple such that a soliton may jump from one site to a neighboring site without coupling to the non-localized excited states.

As the soliton does not couple to the legs, the legs do not couple to the soliton.
Thus, we have shown that when a particle that is initially confined to one leg is allowed to propagate onto the other legs, its dynamics are almost solely described by the four lowest excited states that belong to the same approximately degenerate particle-in-a-box multiplet.
By imposing an external field upon the well or capturing a second particle in the well, we might be able to control the behavior of the primary particle and guide its propagation through the well.
This could open opportunities to design X-well-based transistors for general quantum networks.


Finally, we have considered the possibility of an effective one-dimensional quantum-graph description of the X well.
Our results suggest that, in the extreme limit of infinitesimally thin legs, the legs decouple for all wave numbers above threshold.
Based on our work, it is highly doubtful whether a quantum graph model is useful in describing the dynamics of a physical X well.

Throughout our analysis, we have assumed that the X well is build from flat wires whose boundaries are infinite potential barriers.
Though we do not find these assumptions to be unreasonable on physical grounds, it could be an interesting extension to study the applicability of our results upon lifting these assumptions.
We anticipate that the qualitative features of our results will not change much if we change the geometry of the wires to have, e.g., a square or circular cross section; see also \textcite{delitsyn2012}.
Likewise, X-well whose legs are not at right angles to each other have been considered in \textcite{bulgakov2002}; see also the 
review in the introduction of \textcite{exner2017}. Our formalism could be adapted to such cases as well, but it would require a careful reconsideration of how one defines the modes in order to retain the physical picture and corresponding intuitions that is obtained for the perpendicular crossings studied here.
Networks for which the boundaries are finite potential barrier that allow for evanescent waves outside the wires are studied in the theory of `leaky' quantum graphs reviewed by \textcite{exner2008}. This is analogous to interesting recent experimental development in photonic nanostructures and nanofibers where the evanescent waves of the light field is made to interact with near-by atoms \cite{scheucher2016}.

The intriguing question of multi-particle states in the networks, and the presence of interactions in such systems 
has been touched upon in previous quantum graph approaches \cite{melnikov1995,bolte2013}. There is also the mentioned work on cold atomic condensates in wave guides \cite{markowsky2014} using mean-field theory and the resulting non-linear Schr{\"o}dinger equation. 
However, it does not appear that 
this problem has been considered in great detail starting from just a few interacting particles 
in the geometry.
In our physical approach, we can address multi-particle
systems of non-interacting particles rather easily since we have access to eigenstates. Including interactions 
through a perturbative approach would therefore be straightforward and an interesting topic of future investigation.

%

\begin{acknowledgments}
The results reported here are partly based on the Master's thesis of M.E.S.A. 
The authors would like to thank Manuel Valiente, Signe Thorsen, Kristian Nielsen and Jan Philip Solovej for discussion during various 
stages. This work was support in part by the Danish Council for Independent Research DFF and the
Sapere Aude program, as well as by the Carlsberg Foundation through the Distinguished Associate Professor
Fellowship.
\end{acknowledgments}

\appendix
\onecolumngrid

\section{Matrix elements of the symmetric X well}
\label{sec:symmetric-xwell-matrix-elements}

In this appendix, we compute the matrix elements of $\Psi$ and $\Pi$ for the symmetric X well.

\subsection{Matrix elements of $\Psi$}

Letting
\begin{equation}
  R_{nm} = \frac{nm}{(n^2+m^2-(ka/\pi)^2)^2} a^2,
\end{equation}
we compute the overlaps between perpendicular legs
\begin{equation}
  \braket{n,\uleg|m,\rleg} = r_+ (-1)^{n+m} R_{nm} \quad\text{and}\quad
  \braket{n,\rleg|m,\dleg} = r_- R_{nm}.
\end{equation}
With Eq.~\eqref{eq:mode-relations}, we then notice that the remaining overlaps are
\begin{gather}
  \braket{n,\lleg|m,\uleg} = \braket{n,\rleg|\sigma_- C_4^2|m,\rleg} = \braket{n,\rleg|\sigma_+|m,\rleg} = \braket{n,\uleg|m,\rleg}, \\
  \braket{n,\dleg|m,\lleg} = \braket{n,\rleg|C_4^2 \sigma_+|m,\rleg} = \braket{n,\rleg|\sigma_-|m,\rleg} = \braket{n,\rleg|m,\dleg}.
\end{gather}
Let $t_m = \kpar a$.
The overlap between an eastern-leg mode and itself is
\begin{multline}
  \braket{m,\rleg|m,\rleg} = \frac{a^2}{4t_m} \Bigg( \csc^2(t_m) \left( t_m - \frac{1}{2}\sin(2t_m) \right) 
  -\cot(t_m(L-\half))+t_m(L-\half)\csc^2(t_m(L-\half)) \Bigg).
\end{multline}
This is the same for the other legs as $\sigma_+^2=\sigma_-^2=E$.
We remark that if $t_m$ is purely imaginary,
\begin{multline}
  -\cot(t_m(L-\half))+t_m(L-\half)\csc^2(t_m(L-\half)) 
  = i\coth(\imag(t_m)(L-\half))-t_m(L-\half)\csch^2(\imag(t_m)(L-\half))
  \simeq i
\end{multline}
for $L\gg1$; so the matrix element is independent of $L$ in that limit.
The central-region overlap between modes from opposite legs with equal mode number $m$ is
\begin{equation}
  \braket{m,\lleg|m,\rleg} = \braket{m,\dleg|m,\uleg} = r_+ r_- (-1)^{m+1} \frac{a^2}{4} \csc(t_m) \left(\frac{1}{t_m} - \cot(t_m)\right).
\end{equation}

In conclusion, the matrix elements of $\Psi$ are
\begin{align}
  \Psi_{nm} = (r_+ (-1)^{n+m} + r_-) R_{nm} + 4\delta_{nm} (\braket{m,\rleg|m,\rleg} + \braket{m,\lleg|m,\rleg}).
\end{align}

\subsection{Matrix elements of $\Pi$}

In the following, we find an expression for the energy contribution due to kinks at the interface between the central region and the eastern leg.
By symmetry, the other three interfaces each give the same contribution.

The matrix elements of $\Pi$ are given by
\begin{equation}
  \Pi_{nm} = 4 \sum_{s} \int_{-\half a}^{\half a} \dif{y}\; \sin(k_{n\perp}(\half a+y)) \, \Delta\!\left(\pd{}{x}\braket{x,y|m,s}\right)_{x=\half a}. \label{eq:derivative-matrix-elements}
\end{equation}
The factor of $4$ accounts for the four interfaces.
From the central-region side of the interface, the mode wave functions have derivatives
\begin{align}
  \pd{}{x}\braket{x,y|m,\rleg}\Big\vert_{x\uparrow\half} &= \kpar \cot(\kpar a) \sin(\kper(\half a+y)), \\
  \pd{}{x}\braket{x,y|m,\uleg}\Big\vert_{x\uparrow\half} &= \kper r_+ (-1)^{m} \csc(\kpar a) \sin(\kpar(\half a+y)), \\
  \pd{}{x}\braket{x,y|m,\lleg}\Big\vert_{x\uparrow\half} &= -\kpar r_+ r_- \csc(\kpar a) \sin(\kper(\half a-y)), \\
  \pd{}{x}\braket{x,y|m,\dleg}\Big\vert_{x\uparrow\half} &= -\kper r_- \csc(\kpar a) \sin(\kpar(\half a-y)).
\end{align}
Meanwhile, the derivative approaching from the leg side is
\begin{equation}
  \pd{}{x}\braket{x,y|m,\rleg}\Big\vert_{x\downarrow\half} = -\kpar \cot(\kpar a(L-\half))\sin(\kper(\half a+y)).
\end{equation}
Using that $\sin(\kper(\half a-y))=(-1)^{m+1}\sin(\kper(\half a+y))$, the change in derivative over the interface is
\begin{multline}
  \Delta\!\left(\sum_s \pd{}{x}\braket{x,y|m,s}\right)_{x=\half}
  = -\kpar \Big( \cot(\kpar a(L-\half)) + \cot(\kpar a) \\
    + r_+ r_- (-1)^m \csc(\kpar a) \Big)
  - \kper\csc(\kpar a) \Big( r_+ (-1)^m \sin(\kpar(\half a+y)) 
    - r_- \sin(\kpar(\half a-y)) \Big).
\end{multline}
Hence, the entries of $\Pi$ are
\begin{multline}
  \Pi_{nm}
  = 4(r_+ (-1)^{n+m} + r_-) \frac{nm}{n^2+m^2 - (ka/\pi)^2} 
  - 2t_m \left( \cot(t_m(L-\half)) + \cot(t_m) + r_+ r_- (-1)^m \csc(t_m) \right) \delta_{nm}. \label{eq:Pi-matrix-elements}
\end{multline}

\twocolumngrid

\end{document}